\documentstyle[10pt,epsfig,dp_delphititle,float,hangcaption,xspace,amssymb,
amsfonts,amsmath,amsthm,cite,graphicx]{dp_delphi}
\makeindex
\def\DpPaperGroup{PH--EP}
\def\DpPaperRef{2010--056}
\def\DpDate{9 September 2010}
\def\DpAuthors{DELPHI Collaboration}
\def\DpSubmit{(Accepted by Eur. Phys. J. C)}
\def\DpTitle{
  Search for single top quark production via contact interactions at LEP2
}

\newcommand{\NIM}[3]{{\em Nucl. Instrum. Methods} {\bf{#1} }{(#2) }{#3}}
\newcommand{\NPB}[3]{{\em Nucl. Phys.} {\bf B}{\bf{#1} }{(#2) }{#3}}
\newcommand{\PLB}[3]{{\em Phys. Lett.} {\bf B}{\bf{#1} }{(#2) }{#3}}

\newcommand{\PRD}[3]{{\em Phys. Rev.} {\bf D}{\bf{#1} }{(#2) }{#3}}

\newcommand{\CPC}[3]{{\em Comp. Phys. Comm.} {\bf{#1} }{(#2) }{#3}}
\newcommand{\EJC}[3]{{\em Eur. Phys. J.} {\bf C}{\bf{#1} }{(#2) }{#3}}

\begin{document}
\makeatletter
\newcount\@tempcntc
\def\@citex[#1]#2{\if@filesw\immediate\write\@auxout{\string\citation{#2}}\fi
  \@tempcnta\z@\@tempcntb\m@ne\def\@citea{}\@cite{\@for\@citeb:=#2\do
    {\@ifundefined
       {b@\@citeb}{\@citeo\@tempcntb\m@ne\@citea\def\@citea{,}{\bf ?}\@warning
       {Citation `\@citeb' on page \thepage \space undefined}}%
    {\setbox\z@\hbox{\global\@tempcntc0\csname b@\@citeb\endcsname\relax}%
     \ifnum\@tempcntc=\z@ \@citeo\@tempcntb\m@ne
       \@citea\def\@citea{,}\hbox{\csname b@\@citeb\endcsname}%
     \else
      \advance\@tempcntb\@ne
      \ifnum\@tempcntb=\@tempcntc
      \else\advance\@tempcntb\m@ne\@citeo
      \@tempcnta\@tempcntc\@tempcntb\@tempcntc\fi\fi}}\@citeo}{#1}}
\def\@citeo{\ifnum\@tempcnta>\@tempcntb\else\@citea\def\@citea{,}%
  \ifnum\@tempcnta=\@tempcntb\the\@tempcnta\else
   {\advance\@tempcnta\@ne\ifnum\@tempcnta=\@tempcntb \else \def\@citea{--}\fi
    \advance\@tempcnta\m@ne\the\@tempcnta\@citea\the\@tempcntb}\fi\fi}
 
\makeatother

\begin{titlepage}
\pagenumbering{roman}

\CERNpreprint{\DpPaperGroup}{\DpPaperRef}   
\date{{\small\DpDate}}                      
\title{\DpTitle}                            
\address{\DpAuthors}                        

\begin{shortabs}                            
\noindent
Single top quark production via four-fermion contact interactions
associated to flavour-changing neutral currents was searched for in
data taken by the DELPHI detector at LEP2. The data were
accumulated at centre-of\discretionary{-}{-}{-}mass energies ranging from
189 to 209 GeV, with an integrated luminosity of 598.1~pb$^{-1}$.  No
evidence for a signal was found. Limits on the energy scale $\Lambda$, were
set for scalar-, vector- and tensor-like coupling scenarios.
\end{shortabs}

\vfill

\begin{center}
\DpSubmit \ \\          
\end{center}

\vfill
\clearpage

\headsep 10.0pt

\addtolength{\textheight}{10mm}
\addtolength{\footskip}{-5mm}
\begingroup
%
\newcommand{\DpName}[2]{\hbox{#1$^{\ref{#2}}$},\hfill}
\newcommand{\DpNameTwo}[3]{\hbox{#1$^{\ref{#2},\ref{#3}}$},\hfill}
\newcommand{\DpNameThree}[4]{\hbox{#1$^{\ref{#2},\ref{#3},\ref{#4}}$},\hfill}
\newskip\Bigfill \Bigfill = 0pt plus 1000fill
\newcommand{\DpNameLast}[2]{\hbox{#1$^{\ref{#2}}$}\hspace{\Bigfill}}

%
\footnotesize
\noindent
\DpName{J.Abdallah}{LPNHE}
\DpName{P.Abreu}{LIP}
\DpName{W.Adam}{VIENNA}
\DpName{P.Adzic}{DEMOKRITOS}
\DpName{T.Albrecht}{KARLSRUHE}
\DpName{R.Alemany-Fernandez}{CERN}
\DpName{T.Allmendinger}{KARLSRUHE}
\DpName{P.P.Allport}{LIVERPOOL}
\DpName{U.Amaldi}{MILANO2}
\DpName{N.Amapane}{TORINO}
% ...
\normalsize
\endgroup
\newpage
\titlefoot{Department of Physics and Astronomy, Iowa State
     University, Ames IA 50011-3160, USA
    \label{AMES}}
\titlefoot{Physics Department, Universiteit Antwerpen,
     Universiteitsplein 1, B-2610 Antwerpen, Belgium
    \label{ANTWERP}}
\titlefoot{IIHE, ULB-VUB,
     Pleinlaan 2, B-1050 Brussels, Belgium
    \label{BRUSSELS}}
\titlefoot{Physics Laboratory, University of Athens, Solonos Str.
     104, GR-10680 Athens, Greece
    \label{ATHENS}}
\titlefoot{Department of Physics, University of Bergen,
     All\'egaten 55, NO-5007 Bergen, Norway
    \label{BERGEN}}
\titlefoot{Dipartimento di Fisica, Universit\`a di Bologna and INFN,
     Viale C. Berti Pichat 6/2, IT-40127 Bologna, Italy
    \label{BOLOGNA}}
\titlefoot{Centro Brasileiro de Pesquisas F\'{\i}sicas, rua Xavier Sigaud 150,
     BR-22290 Rio de Janeiro, Brazil
    \label{BRASIL-CBPF}}
\titlefoot{Inst. de F\'{\i}sica, Univ. Estadual do Rio de Janeiro,
     rua S\~{a}o Francisco Xavier 524, Rio de Janeiro, Brazil
    \label{BRASIL-IFUERJ}}
\titlefoot{Coll\`ege de France, Lab. de Physique Corpusculaire, IN2P3-CNRS,
     FR-75231 Paris Cedex 05, France
    \label{CDF}}
\titlefoot{CERN, CH-1211 Geneva 23, Switzerland
    \label{CERN}}
\titlefoot{Institut Pluridisciplinaire Hubert Curien, Universit\'e de Strasbourg,
     IN2P3-CNRS, BP28, FR-67037 Strasbourg \indent~~Cedex~2, France
    \label{CRN}}
\titlefoot{Now at DESY-Zeuthen, Platanenallee 6, D-15735 Zeuthen, Germany
    \label{DESY}}
\titlefoot{Institute of Nuclear Physics, N.C.S.R. Demokritos,
     P.O. Box 60228, GR-15310 Athens, Greece
    \label{DEMOKRITOS}}
\titlefoot{FZU, Inst. of Phys. of the C.A.S. High Energy Physics Division,
     Na Slovance 2, CZ-182 21, Praha 8, Czech Republic
    \label{FZU}}
\titlefoot{Dipartimento di Fisica, Universit\`a di Genova and INFN,
     Via Dodecaneso 33, IT-16146 Genova, Italy
    \label{GENOVA}}
\titlefoot{Institut des Sciences Nucl\'eaires, IN2P3-CNRS, Universit\'e
     de Grenoble 1, FR-38026 Grenoble Cedex, France
    \label{GRENOBLE}}
\titlefoot{Helsinki Institute of Physics and Department of Physical Sciences,
     P.O. Box 64, FIN-00014 University of Helsinki, 
     \indent~~Finland
    \label{HELSINKI}}
\titlefoot{Joint Institute for Nuclear Research, Dubna, Head Post
     Office, P.O. Box 79, RU-101 000 Moscow, Russian Federation
    \label{JINR}}
\titlefoot{Institut f\"ur Experimentelle Kernphysik,
     Universit\"at Karlsruhe, Postfach 6980, DE-76128 Karlsruhe,
     Germany
    \label{KARLSRUHE}}
\titlefoot{Institute of Nuclear Physics PAN,Ul. Radzikowskiego 152,
     PL-31142 Krakow, Poland
    \label{KRAKOW1}}
\titlefoot{Faculty of Physics and Nuclear Techniques, University of Mining
     and Metallurgy, PL-30055 Krakow, Poland
    \label{KRAKOW2}}
\titlefoot{LAL, Univ Paris-Sud, CNRS/IN2P3, Orsay, France
    \label{LAL}}
\titlefoot{School of Physics and Chemistry, University of Lancaster,
     Lancaster LA1 4YB, UK
    \label{LANCASTER}}
\titlefoot{LIP, FCUL, IST, CFC-UC - Av. Elias Garcia, 14-$1^{o}$,
     PT-1000 Lisboa Codex, Portugal
    \label{LIP}}
\titlefoot{Department of Physics, University of Liverpool, P.O.
     Box 147, Liverpool L69 3BX, UK
    \label{LIVERPOOL}}
\titlefoot{Dept. of Physics and Astronomy, Kelvin Building,
     University of Glasgow, Glasgow G12 8QQ, UK
    \label{GLASGOW}}
\titlefoot{LPNHE, IN2P3-CNRS, Univ.~Paris VI et VII, Tour 33 (RdC),
     4 place Jussieu, FR-75252 Paris Cedex 05, France
    \label{LPNHE}}
\titlefoot{Department of Physics, University of Lund,
     S\"olvegatan 14, SE-223 63 Lund, Sweden
    \label{LUND}}
\titlefoot{Universit\'e Claude Bernard de Lyon, IPNL, IN2P3-CNRS,
     FR-69622 Villeurbanne Cedex, France
    \label{LYON}}
\titlefoot{Dipartimento di Fisica, Universit\`a di Milano and INFN-MILANO,
     Via Celoria 16, IT-20133 Milan, Italy
    \label{MILANO}}
\titlefoot{Dipartimento di Fisica, Univ. di Milano-Bicocca and
     INFN-MILANO, Piazza della Scienza 3, IT-20126 Milan, Italy
    \label{MILANO2}}
\titlefoot{IPNP of MFF, Charles Univ., Areal MFF,
     V Holesovickach 2, CZ-180 00, Praha 8, Czech Republic
    \label{NC}}
\titlefoot{NIKHEF, Postbus 41882, NL-1009 DB
     Amsterdam, The Netherlands
    \label{NIKHEF}}
\titlefoot{National Technical University, Physics Department,
     Zografou Campus, GR-15773 Athens, Greece
    \label{NTU-ATHENS}}
\titlefoot{Physics Department, University of Oslo, Blindern,
     NO-0316 Oslo, Norway
    \label{OSLO}}
\titlefoot{Dpto. Fisica, Univ. Oviedo, Avda. Calvo Sotelo
     s/n, ES-33007 Oviedo, Spain
    \label{OVIEDO}}
\titlefoot{Department of Physics, University of Oxford,
     Keble Road, Oxford OX1 3RH, UK
    \label{OXFORD}}
\titlefoot{Dipartimento di Fisica, Universit\`a di Padova and
     INFN, Via Marzolo 8, IT-35131 Padua, Italy
    \label{PADOVA}}
\titlefoot{Rutherford Appleton Laboratory, Chilton, Didcot
     OX11 OQX, UK
    \label{RAL}}
\titlefoot{Dipartimento di Fisica, Universit\`a di Roma II and
     INFN, Tor Vergata, IT-00173 Rome, Italy
    \label{ROMA2}}
\titlefoot{Dipartimento di Fisica, Universit\`a di Roma III and
     INFN, Via della Vasca Navale 84, IT-00146 Rome, Italy
    \label{ROMA3}}
\titlefoot{DAPNIA/Service de Physique des Particules,
     CEA-Saclay, FR-91191 Gif-sur-Yvette Cedex, France
    \label{SACLAY}}
\titlefoot{Instituto de Fisica de Cantabria (CSIC-UC), Avda.
     los Castros s/n, ES-39006 Santander, Spain
    \label{SANTANDER}}
\titlefoot{Inst. for High Energy Physics, Serpukov
     P.O. Box 35, Protvino, (Moscow Region), Russian Federation
    \label{SERPUKHOV}}
\titlefoot{J. Stefan Institute, Jamova 39, SI-1000 Ljubljana, Slovenia
    \label{SLOVENIJA1}}
\titlefoot{Laboratory for Astroparticle Physics,
     University of Nova Gorica, Kostanjeviska 16a, SI-5000 Nova Gorica, Slovenia
    \label{SLOVENIJA2}}
\titlefoot{Department of Physics, University of Ljubljana,
     SI-1000 Ljubljana, Slovenia
    \label{SLOVENIJA3}}
\titlefoot{Fysikum, Stockholm University,
     Box 6730, SE-113 85 Stockholm, Sweden
    \label{STOCKHOLM}}
\titlefoot{Dipartimento di Fisica Sperimentale, Universit\`a di
     Torino and INFN, Via P. Giuria 1, IT-10125 Turin, Italy
    \label{TORINO}}
\titlefoot{INFN,Sezione di Torino and Dipartimento di Fisica Teorica,
     Universit\`a di Torino, Via Giuria 1,
     IT-10125 Turin, Italy
    \label{TORINOTH}}
\titlefoot{Dipartimento di Fisica, Universit\`a di Trieste and
     INFN, Via A. Valerio 2, IT-34127 Trieste, Italy
    \label{TRIESTE}}
\titlefoot{Istituto di Fisica, Universit\`a di Udine and INFN,
     IT-33100 Udine, Italy
    \label{UDINE}}
\titlefoot{Univ. Federal do Rio de Janeiro, C.P. 68528
     Cidade Univ., Ilha do Fund\~ao
     BR-21945-970 Rio de Janeiro, Brazil
    \label{UFRJ}}
\titlefoot{Department of Radiation Sciences, University of
     Uppsala, P.O. Box 535, SE-751 21 Uppsala, Sweden
    \label{UPPSALA}}
\titlefoot{IFIC, Valencia-CSIC, and D.F.A.M.N., U. de Valencia,
     Avda. Dr. Moliner 50, ES-46100 Burjassot (Valencia), Spain
    \label{VALENCIA}}
\titlefoot{Institut f\"ur Hochenergiephysik, \"Osterr. Akad.
     d. Wissensch., Nikolsdorfergasse 18, AT-1050 Vienna, Austria
    \label{VIENNA}}
\titlefoot{Inst. Nuclear Studies and University of Warsaw, Ul.
     Hoza 69, PL-00681 Warsaw, Poland
    \label{WARSZAWA}}
\titlefoot{Now at University of Warwick, Coventry CV4 7AL, UK
    \label{WARWICK}}
\titlefoot{Fachbereich Physik, University of Wuppertal, Postfach
     100 127, DE-42097 Wuppertal, Germany \\
\noindent
{$^\dagger$~deceased}
    \label{WUPPERTAL}}
\addtolength{\textheight}{-10mm}
\addtolength{\footskip}{5mm}
\clearpage

\headsep 30.0pt
\end{titlepage}

%
\pagenumbering{arabic}                              
\setcounter{footnote}{0}                            %
\large
\newcommand{\Xb}{{\mathrm{b}}}
\newcommand{\Xc}{{\mathrm{c}}}
\newcommand{\Xu}{{\mathrm{u}}}
\newcommand{\Xt}{{\mathrm{t}}}
\newcommand{\Xq}{{\mathrm{q}}}
\newcommand{\Xj}{{\mathrm{j}}}
\newcommand{\Xf}{{\mathrm{f}}}
\newcommand{\Xe}{{\mathrm{e}}}
\newcommand{\XW}{{\mathrm{W}}}
\newcommand{\XZ}{{\mathrm{Z}}}
\newcommand{\Xo}{\phantom{.0}}


\section{Introduction}

With a mass of $173.3\pm 0.6\pm 0.9$~GeV\cite{tevatronmass}, the $\Xt$ quark is
the heaviest known one and, due to its large mass, the most sensitive to
new physics. In $\Xe^+\Xe^-$ collisions at LEP2, $\Xt$ quarks could only
be singly produced, due to the limited centre-of-mass energy. In the
Standard Model (SM) they would be generated in association with $\Xb$ or
$\Xc$ quarks, through the processes\footnote{Throughout this paper the
charge conjugated processes are also included, unless explicitly stated
otherwise.} $\Xe^+\Xe^-\to \Xt\bar \Xb\Xe^-\bar{\nu_\Xe}$ and
$\Xe^+\Xe^-\to \Xt\bar \Xc$. A complete tree level calculation has shown
that the cross-section of the first process is at the level of
$10^{-6}$~pb~\cite{boos}. The second process proceeds via Flavour
Changing Neutral Currents~(FCNC), which are forbidden at tree level and
suppressed by the GIM mechanism~\cite{gim} at higher orders. The
corresponding cross-section is of the order of
$10^{-12}$~pb~\cite{huang}. 

Enhanced $\Xe^+\Xe^-\to \Xt\bar \Xc$ cross-sections (or top FCNC
branching ratios) are, however, foreseen in several new physics
scenarios, such as models with extra $Q=2/3$ quark singlets~\cite{juan},
two Higgs doublet models~(2HDM)~\cite{atwood1,atwood2}, flavour
conserving 2HDM~\cite{atwood2,bejar}, minimal supersymmetric
SM~\cite{liu,delepine,guasch} or non-minimal supersymmetric models with
$R$ parity violation~\cite{yang}. Single $\Xt$ quark production at LEP2
would thus be a signature of new physics.

The four LEP collaborations~\cite{aleph00,aleph02,fcnc,l3,opal} searched
previously for single $\Xt$ production in the context of
Ref.~\cite{obraztsov}. In the model used, single $\Xt$ production is
described in terms of vector-like anomalous couplings ($\kappa_\XZ$ and
$\kappa_\gamma$) associated with the already known $\XZ$ and $\gamma$
bosons. The physics energy scale was set to the $\Xt$ mass.

In this paper, a very general approach describing single $\Xt$ quark
production via $\Xe^+\Xe^-\to \Xt\bar \Xc$ through an effective
Lagrangian with FCNC operators is used~\cite{wudka}. Apart from the $\XZ$ and
$\gamma$ bosons, new four-fermion contact interactions, which include
additional scalar-, vector-, and tensor-like couplings, are possible. 
The contribution of the $\XZ$ boson is also included, through a
vector-like coupling which can be related to the anomalous coupling
$\kappa_\XZ$~\cite{obraztsov}. The physics energy scale is a free parameter in this
model. The kinematic differences between different coupling assumptions
are taken into account and a dedicated analysis is developed.

This paper is organized as follows: single t quark production and decay
is briefly discussed in Section~\ref{sec:production}. In
Section~\ref{sec:data} the data sets and the simulated samples are
presented. The analysis is described in Section~\ref{sec:analysis} and
the results are presented in Section~\ref{sec:results}. In
Section~\ref{sec:conclusions}, conclusions are drawn and the results are
compared with previous LEP studies.


\section{Single t quark production and decay\label{sec:production}}

\begin{figure}[p]
  \begin{center}
    \includegraphics[width=.6\textwidth]{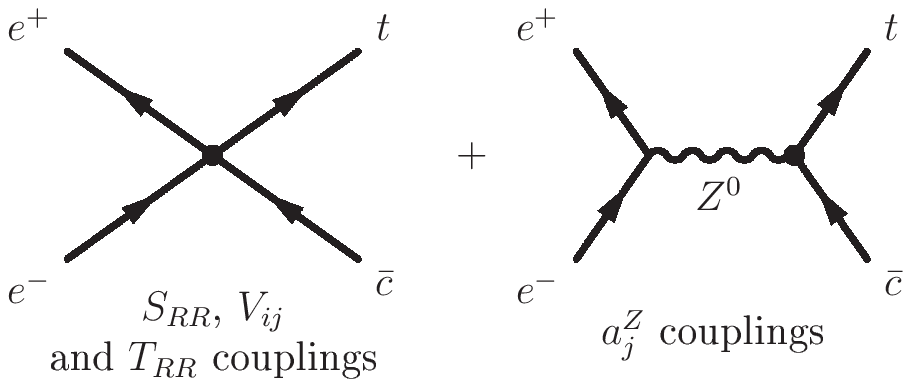}
  \end{center}
  \caption{The $\Xe\Xe\Xt\Xc$ and $\XZ\Xt\Xc$ vertex contributions to the 
$\Xe^+\Xe^-\to \Xt\bar \Xc$ process.
  \label{fig:feynman}}
\end{figure}

The process $\Xe^+\Xe^-\to \Xt\bar \Xc$ can be described by an effective
Lagrangian with FCNC operators~\cite{wudka}. Fig.~\ref{fig:feynman}
shows the Feynman diagrams considered in this model. New contact
interaction terms are associated to new scalar ($S_{RR}$), vector
($V_{ij}$, $i,j=L,R$) and tensor-like ($T_{RR}$) couplings of heavy
fields, and a term describing a new $\XZ\Xt\Xc$ vertex parametrized by vector
($a^Z_j$) couplings is introduced.

\begin{figure}[p]
  \begin{center}
    \includegraphics[width=.96\textwidth]{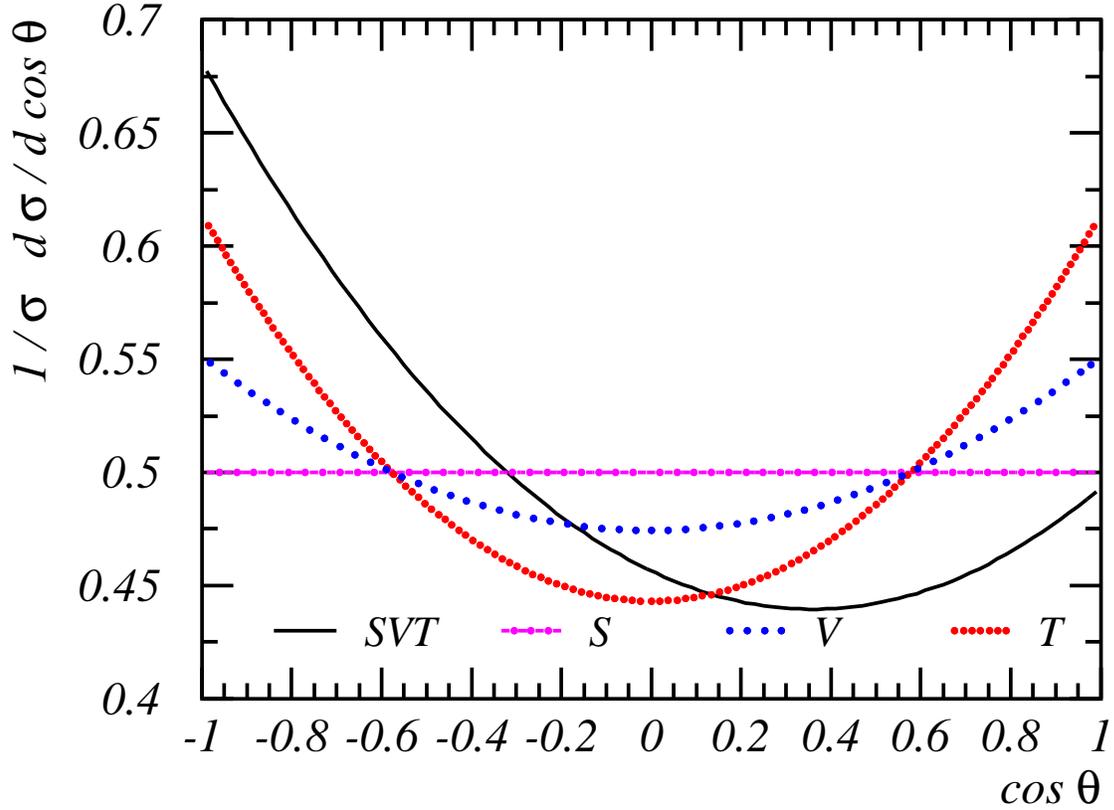}
  \end{center}
  \caption{The differential cross-section 
  $\mathrm{d}\sigma/\mathrm{d}\cos\theta$, normalized to the total
  cross-section, for the process $\Xe^+\Xe^-\to \Xt\bar \Xc$ without ISR, is
  shown as a function of the cosine of the polar angle of the $\Xt$ quark, 
  for $m_\Xt=175$~GeV$/c^2$, $\Lambda=1$~TeV, $\sqrt{s}=206$~GeV and the 
  scenarios described in Table~\ref{tab:scenarios}. The shapes of the 
  differential cross-sections for scenarios $a$, $V-a$ and $V+a$ are the same 
  as scenario $V$.
  \label{fig:diff.cross.section}}
\end{figure}

The differential cross-section, for $\Xt\bar \Xc$ production only, can
be obtained from the Lagrangian given in Ref.~\cite{wudka} and is
expressed in terms of the couplings and of a new physics energy scale
parameter $\Lambda$:
{\allowdisplaybreaks
\begin{eqnarray}\label{eq:dsig}
  \frac{\mathrm{d}\sigma}{\mathrm{d}\cos\theta}(\Xe^+\Xe^-\to \Xt\bar \Xc)=%
  \frac{3\mathcal{C}}{8}\Biggl\{
  &\hspace{-8mm}&  
    S^2_{RR}
    (1+\beta)
    -4S_{RR}T_{RR}
    (1+\beta)\cos\theta
  \phantom{\Biggl\{}\nonumber\\&\hspace{-8mm}&
    +16T^2_{RR}
    \left(1-\beta+2\beta\cos^2\theta\right)
  \phantom{\Biggl\{}\nonumber\\&\hspace{-8mm}&
    +2\left[
      \left(V_{RR} + 4 c_R^Z a_R^Z\frac{m_Wm_Z}{s-m^2_Z}\right)^2+
      \left(V_{LL} + 4 c_L^Z a_L^Z\frac{m_Wm_Z}{s-m^2_Z}\right)^2
    \right]
  \phantom{\Biggl\{}\nonumber\\&\hspace{-8mm}&
    \quad\times\left[1+(1+\beta)\cos\theta+\beta\cos^2\theta\right]
  \phantom{\Biggl\{}\nonumber\\&\hspace{-8mm}&
    +2\left[
      \left(V_{RL} + 4 c_R^Z a_L^Z\frac{m_Wm_Z}{s-m^2_Z}\right)^2+
      \left(V_{LR} + 4 c_L^Z a_R^Z\frac{m_Wm_Z}{s-m^2_Z}\right)^2
    \right]
  \phantom{\Biggl\{}\nonumber\\&\hspace{-8mm}&
    \quad\times\left[1-(1+\beta)\cos\theta+\beta\cos^2\theta\right]
  \Biggr\},
\end{eqnarray}}
where 
\begin{displaymath}
  {\mathcal{C}}=\frac{s}{\Lambda^4}\frac{\beta^2}{4\pi(1+\beta)^3},\quad
  \beta=\frac{(s-m_\Xt^2)}{(s+m_\Xt^2)},\quad
  c_L^Z = -1/2 + \sin^2\theta_W,\quad
  c^Z_R=\sin^2\theta_W,
\end{displaymath}
$\beta$ is the velocity of the outgoing $\Xt$ quark, $\theta_\XW$ is the
electroweak mixing angle and $\theta$ is the angle between the incident
electron beam and the $\Xt$ quark. The coupling scenarios listed in
Table~\ref{tab:scenarios} were considered in this study. The ``$V+a$''
(``$V-a$'') scenario corresponds to the constructive (destructive)
interference between the $\Xe\Xe\Xt\Xc$ and the $\XZ\Xt\Xc$ vertices.
The differential cross-section depends on the coupling scenarios as can
be seen in Fig.~\ref{fig:diff.cross.section} for scenarios $SVT$, $S$,
$V$ and $T$, considering $m_\Xt=175$~GeV$/c^2$, $\Lambda=1$~TeV and
$\sqrt{s}=206$~GeV.

\begin{table}[bt]
  \begin{center}\begin{tabular*}{\textwidth}{@{\extracolsep{\fill}}ccccccccc}
    \hline
    \hline
    Scenarios & $S_{RR}$ & $V_{RR}$ & $V_{RL}$ & $V_{LR}$ & $V_{LL}$ & $T_{RR}$ & $a^Z_R$ & $a^Z_L$ \\
    \hline
    \hline
     $SVT$  &    1     &     1    &    1     &     1    &     1    &    1     &    0    &    0     \\
    \hline 
      $S$   &    1     &     0    &    0     &     0    &     0    &    0     &    0    &    0     \\
    \hline
      $V$   &    0     &     1    &    1     &     1    &     1    &    0     &    0    &    0     \\
    \hline
      $T$   &    0     &     0    &    0     &     0    &     0    &    1     &    0    &    0     \\
    \hline
      $a$   &    0     &     0    &    0     &     0    &     0    &    0     &    1    &    1     \\
    \hline
     $V-a$  &    0     &     1    &    1     &     1    &     1    &    0     &  $-$1   &  $-$1    \\
    \hline
     $V+a$  &    0     &     1    &    1     &     1    &     1    &    0     &    1     &   1     \\
    \hline
    \hline
  \end{tabular*}\end{center}
  \caption{The couplings used in the different scenarios considered in 
    this paper.
  \label{tab:scenarios}}
\end{table}

The total production cross-section, including charge conjugation,
obtained from Equ.~\ref{eq:dsig} is
{\allowdisplaybreaks
\begin{eqnarray}\label{eq:sig}
  \sigma(\Xe^+\Xe^-\to \Xt\bar\Xc)  +
  \sigma(\Xe^+\Xe^-\to \bar\Xt\Xc) &=&
  {\mathcal{C}} \Biggl\{8 (3-\beta) T_{RR}^2 +
       \frac{3}{2} (1+\beta) S_{RR}^2 + (3+\beta)\times
    \phantom{\Biggl\{}\nonumber\\&&
    \hspace*{-1.5cm}\biggl[
       \biggl(V_{RR} + 4 c_R^Z a_R^Z\frac{m_Wm_Z}{s-m^2_Z}\biggr)^2
      +\biggl(V_{LL} + 4 c_L^Z a_L^Z\frac{m_Wm_Z}{s-m^2_Z}\biggr)^2+
    \phantom{\Biggl\{}\nonumber\\&&
    \hspace*{-1.5cm}\biggl(V_{RL} + 4 c_R^Z a_L^Z\frac{m_Wm_Z}{s-m^2_Z}\biggr)^2
      +\biggl(V_{LR} + 4 c_L^Z a_R^Z\frac{m_Wm_Z}{s-m^2_Z}\biggr)^2
    \biggr]
  \Biggr\}.
\end{eqnarray}}

\begin{figure}[p]
  \begin{center}
    \includegraphics[width=.96\textwidth]{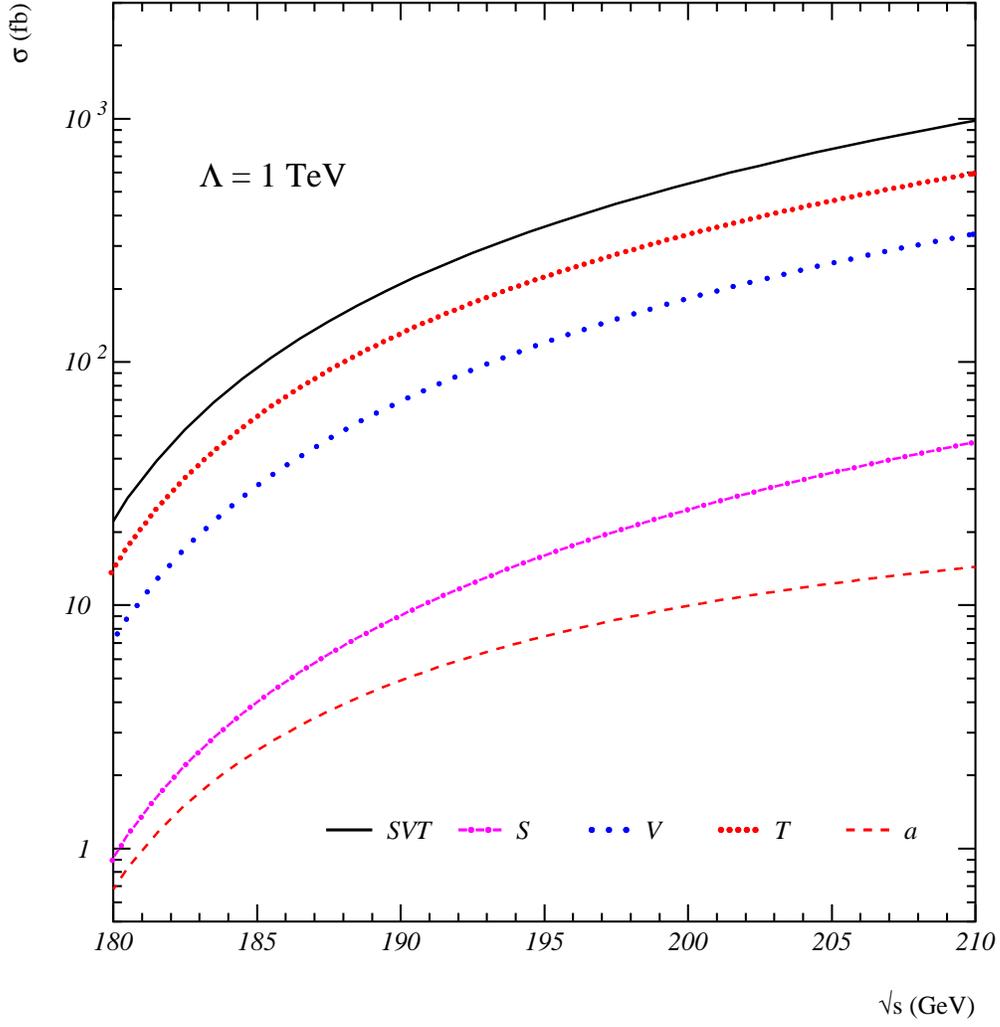}
  \end{center}
  \caption{The total cross-section $\sigma_{\Xt\Xc} = \sigma(\Xe^+ \Xe^- 
  \to\Xt\bar \Xc + \bar \Xt \Xc)$ is shown as a function of the 
  centre-of-mass energy, for $m_\Xt=175$~GeV$/c^2$, $\Lambda=1$~TeV and 
  for the scenarios described in Table~\ref{tab:scenarios}. In this 
  scale the cross-sections for scenarios $V-a$ and $V+a$ are
  indistinguishable from the cross-section for scenario $V$.
  \label{fig:cross.section}}
\end{figure}

The total cross-section as a function of the centre-of-mass energy for
$\Lambda=1$~TeV is represented in Fig.~\ref{fig:cross.section}. It can
be seen that, for the scenarios mentioned above, the contribution from
the $\XZ\Xt\Xc$ vertex is about two orders of magnitude smaller than the
one from the $\Xe\Xe\Xt\Xc$ vertex.

The $\XZ\Xt\Xc$ vertex was described within other models by an anomalous
coupling, $\kappa_\XZ$, as discussed in Ref.~\cite{obraztsov}.
The couplings $\kappa_\XZ$ and $a^\XZ_j$ are related by:
\begin{equation}
  \kappa_\XZ^2 = \left[\left(a^\XZ_L\right)^2 + \left(a^\XZ_R\right)^2\right]
    \left[2 \cos\theta_\XW 
    \left(\frac{v}{\Lambda}\right)^2\right]^2,
  \label{eq:akz}
\end{equation}
where $v=246$~GeV is the SM Higgs vacuum expectation value.

The $\Xt$ quark is expected to decay mainly into $\XW\Xb$. The decays of
the $\XW$ into both quarks and leptons were considered, giving rise to a
hadronic topology ($\Xt\bar \Xc\to \Xb\bar \Xc\Xq\bar {\Xq^\prime}$) and
a semi-leptonic topology ($\Xt\bar \Xc\to \Xb\bar \Xc\ell^+\nu_\ell$).


\section{Data samples and simulation\label{sec:data}}

The data were collected with the DELPHI detector
during the 1998, 1999 and 2000 LEP2 runs at $\sqrt{s}=189-209$~GeV and
correspond to a total integrated luminosity of 598.1~pb${}^{-1}$.
The integrated luminosity collected at each centre-of-mass energy is
shown in Table~\ref{tab:luminosity}.

DELPHI consisted of several sub-detectors in cylindrical layers in
the barrel region and was closed by two endcaps that formed the
forward region of the detector. The main sub-detectors used for the
present analysis were the tracking detectors, the calorimeters and the
muon chambers. Starting from the beam pipe, the barrel tracking
detectors were a three layer silicon micro-vertex detector (the Vertex
Detector), a combined drift/proportional chambers detector (the Inner
Detector), the Time Projection Chamber~(TPC) which was the main tracking
detector and, finally, a 5 layer drift tube detector (the Outer
Detector). The forward region was covered by drift chambers (the Forward
Chambers A and B). The electromagnetic calorimeters were a sampling
calorimeter of lead and gas in the barrel zone, the High-Density
Projection Chamber, and a lead-glass calorimeter with 4532 blocks in
each endcap, the Forward Electromagnetic Calorimeter. The Hadron
Calorimeter was a sampling iron/gas detector in both the barrel and
forward regions, with the iron simultaneously used as the
magnet yoke. The Muon Chambers were sets of drift chambers which formed
the outer surface of the DELPHI detector and were crucial for
identifying muons, essentially the only particles that reached these
detectors. A detailed description of the DELPHI detector can be found in Ref.~\cite{delphi91,delphi96}.
During the year 2000 data taking, an irreversible failure affected one
sector of the TPC, corresponding to 1/12 of its acceptance. The data
recorded under these conditions were analysed separately.

\begin{table}[bt]
  \begin{center}\begin{tabular*}{\textwidth}{@{\extracolsep{\fill}}ccccccccc}
    \hline
    \hline
    year & 1998 & 1999 & 1999 & 1999 & 1999 & 2000 & 2000 & 2000 \\
    \hline
    $\langle\sqrt{s}\rangle$ (GeV) & 188.6 & 191.6 & 195.5 & 199.5 & 201.6 & 204.8 & 206.6 & 206.3$^*$ \\
    \hline
    $\cal L$ (pb$^{-1}$)& 153.0 & 25.1 & 76.0 & 82.7 & 40.2 & 80.0 & 81.9 & 59.2 \\
    \hline
    \hline
  \end{tabular*}\end{center}
  \caption{Integrated luminosity collected with the DELPHI detector at each
  centre-of-mass energy. The data collected during the year 2000 with the 
  TPC fully operational were split into two energy bins, below and above 
  $\sqrt{s}=206$~GeV, with $\langle\sqrt{s}\rangle=204.8$~GeV and 
  \hbox{$\langle\sqrt{s}\rangle=206.6$~GeV}, respectively. The last 
  column, marked by an asterisk, corresponds to data collected with a 
  reduced TPC acceptance.
  \label{tab:luminosity}}
\end{table}

The relevant SM background processes were simulated at each
centre-of-mass energy using several Monte Carlo generators. All the
four-fermion final states (both neutral and charged currents) were
generated with WPHACT~\cite{wphact97,wphact03a,wphact03b}, while the particular phase space
regions of $\Xe^+ \Xe^- \to \Xe^+ \Xe^- \Xf \bar \Xf$ referred to as
$\gamma\gamma$ were generated using PYTHIA~6.1~\cite{pythia61}. The
$\Xq\Xq(\gamma)$ final state was generated with KK2F~\cite{kk2f}.
Processes giving mainly leptonic final states were also generated,
namely Bhabha events with BHWIDE~\cite{bhwide},
$\Xe^+\Xe^-\to\mu^+\mu^-$ events with KK2F and
$\Xe^+\Xe^-\to\tau^+\tau^-$ events with KORALZ~\cite{koralz}. The
fragmentation and hadronisation of the final-state quarks was handled by
PYTHIA~6.1.

For each coupling scenario, signal samples were generated using a modified 
version of PYTHIA~5.7~\cite{pythia57a,pythia57b}, where the angular 
distribution for $\Xt$ quark production was introduced according to 
Equ.~\ref{eq:dsig} and considering $m_\Xt=175$~GeV$/c^2$. Samples with $\Xt$ 
quark masses of 170~GeV and 180~GeV were generated for the evaluation of 
systematic uncertainties. Initial and final state radiation~(ISR and FSR) 
were taken into account. The final-state quarks fragmentation and 
hadronisation was handled by JETSET~7.408~\cite{pythia57a,pythia57b}.

The generated signal and background events were passed through the
detailed simulation of the DELPHI detector~\cite{delphi96} and then
processed with the same reconstruction and analysis programs as the
data.


\section{Analysis description\label{sec:analysis}}

The analysis consisted of a sequential selection used to identify the
event topology and reduce SM background contamination, followed by a
probabilistic analysis based on the construction of a discriminant
variable. With the exception of a common preselection, the hadronic ($\Xt\bar \Xc\to
\Xb\bar \Xc\Xq\bar {\Xq^\prime}$) and the semi-leptonic ($\Xt\bar \Xc\to
\Xb\bar \Xc\ell^+\nu_\ell$) topologies were considered separately at
each step of the analysis.


\subsection{Sequential selection}

A common preselection was adopted for both topologies, followed by 
specific selection criteria.

Events were preselected requiring at least eight good charged-particles
tracks and a visible energy greater than $0.2\sqrt{s}$, measured at
polar angles\footnote{In the standard DELPHI coordinate system, the
positive $z$ axis is along the electron direction. The polar angle
$\theta$ is defined with respect to the $z$ axis. In this paper, polar
angle ranges are always assumed to be symmetric with respect to
$\theta=90^\circ$.} above $20^\circ$. Good charged-particles tracks were
selected by requiring a momentum above 0.2~GeV$/c$ with a relative error
below 1, and impact parameters along the beam direction and in the
transverse plane below 4~cm$/\sin \theta$ and 4~cm, respectively.

The identification of muons relied on the association of
charged particles to signals in the muon chambers and in the hadronic
calorimeter and was provided by standard DELPHI
algorithms~\cite{delphi96}, which classified each identified muon as
\emph{very loose}, \emph{loose}, \emph{standard} or \emph{tight}. The
identification of electrons and photons was performed by combining
information from the electromagnetic calorimeters and the tracking
system. Radiation and interaction effects were taken into account by an
angular clustering procedure around the main shower~\cite{remclu}.
Electron and photon candidates were classified as \emph{loose} or
\emph{tight} by the identification algorithms.

The search for isolated particles (charged leptons and photons) was done by
constructing double cones centered in the direction of charged-particle
tracks or neutral energy deposits. The latter ones were defined as
calorimetric energy deposits above 0.5~GeV, not matched to
charged-particles tracks and identified as photon candidates by the
standard DELPHI algorithms~\cite{delphi96,remclu}. For charged leptons
(photons), the energy in the region between the two cones,
which had half-opening angles of $5^\circ$ and $25^\circ$ ($5^\circ$ and
$15^\circ$), was required to be below 3~GeV (1~GeV),
to ensure isolation. All the charged-particle tracks and neutral energy
deposits inside the inner cone were associated to the isolated particle.
The energy of the isolated particle was then re-evaluated as the sum of the
energies inside the inner cone and was required to be above 5~GeV. For 
\emph{tight} electrons, \emph{standard} or \emph{tight} muons or \emph{tight}
photons the above requirements were weakened. In this
case only the external cone was used and its angle $\alpha$ was varied
according to the energy of the lepton (photon) candidate, down to $2^\circ$
for $P_\ell\geq 70$~GeV/$c$ ($3^\circ$ for $P_\gamma\geq 90$~GeV/$c$), with
the allowed energy inside the cone reduced by $\sin\alpha/\sin25^\circ$
($\sin\alpha/\sin15^\circ$).

The topology of each event was defined according to the number of
isolated charged leptons identified in the event: events with no
isolated charged leptons were classified as hadronic while all the other
events were classified as semi-leptonic. Following the fragmentation and
hadronisation, final state quarks were identified as jets. In both
topologies, a $\Xb$ jet identified using the combined $\Xb$-tag
described in Ref.~\cite{btag}, and a low momentum jet from the $\Xc$
quark were expected. The events of the hadronic (semi-leptonic) topology
were forced into four (two) jets\footnote{Isolated charged leptons and
isolated photons were excluded in the jet clustering.}, using the Durham
jet algorithm~\cite{durham}.

After this common preselection specific selection criteria were applied 
to both topologies.


\subsubsection*{Hadronic topology}

\begin{figure}[p]
  \begin{center}
    \huge DELPHI
  \end{center}

  \includegraphics[width=.48\textwidth]{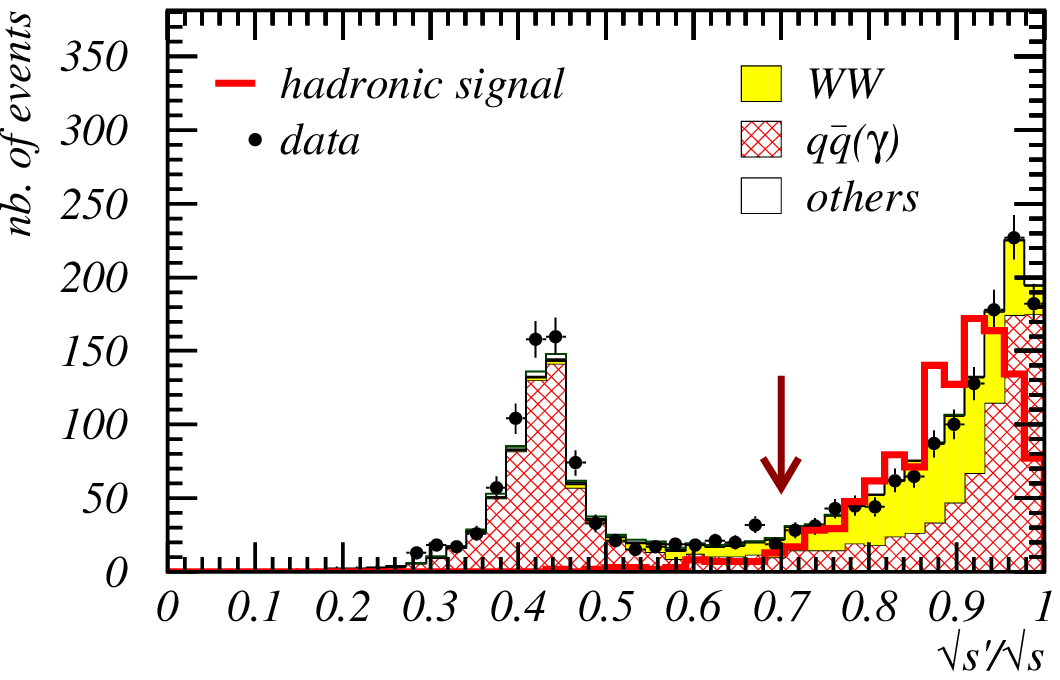}
  \hfill
  \includegraphics[width=.48\textwidth]{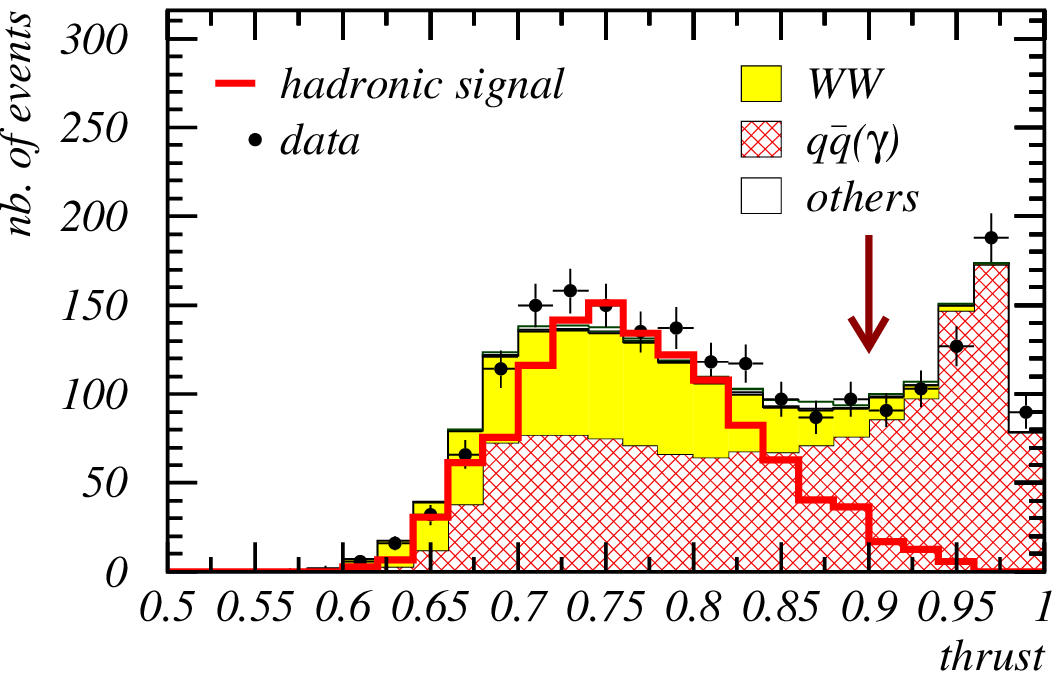}

  \vspace*{-1em}

  \hspace{.24\textwidth}a)\hfill b)\hspace{.24\textwidth}\vspace{1em}

  \includegraphics[width=.48\textwidth]{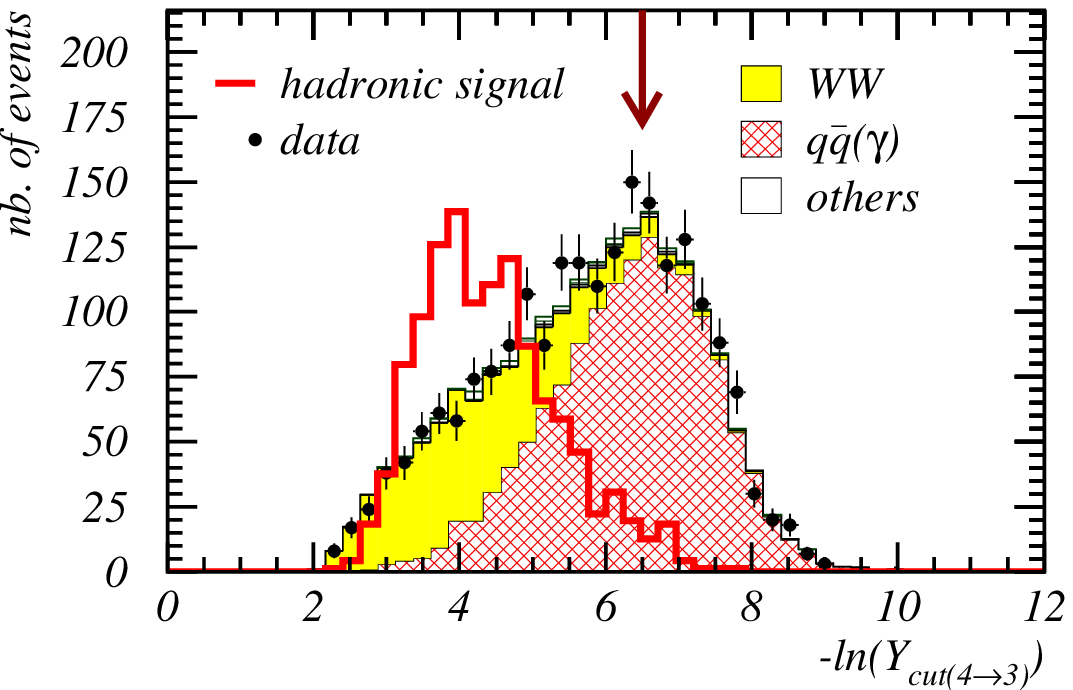}
  \hfill
  \includegraphics[width=.48\textwidth]{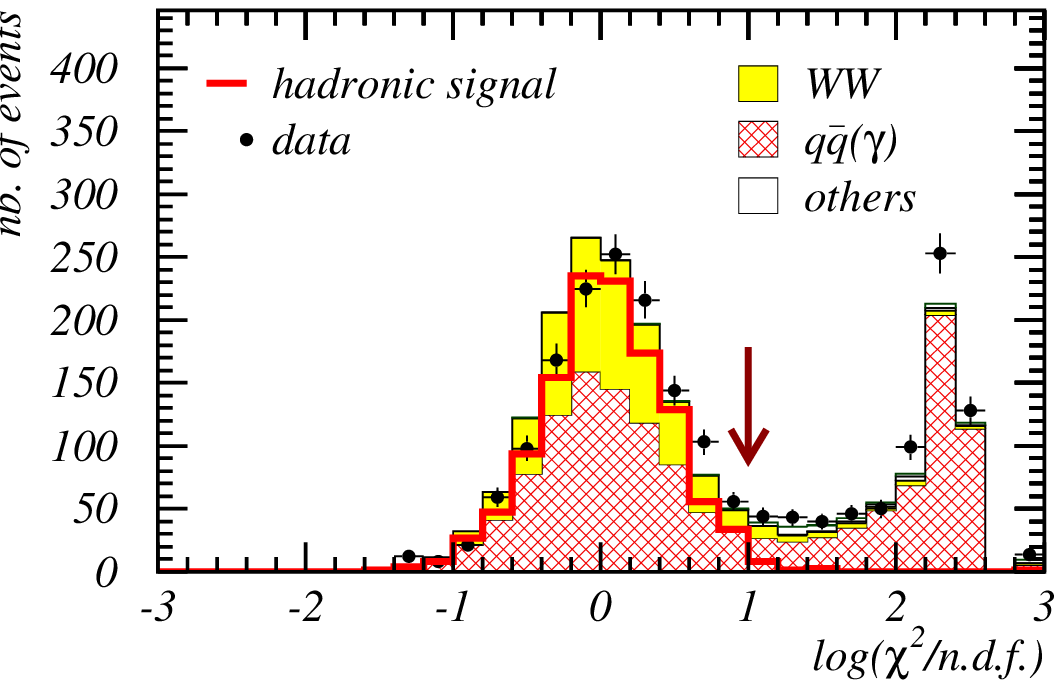}

  \vspace*{-1em}

  \hspace{.24\textwidth}c)\hfill d)\hspace{.24\textwidth}\vspace{1em}

  \caption{Distributions of variables relevant for the sequential 
  selection of the hadronic topology are shown at 
  $\langle\sqrt{s}\rangle=206.6$~GeV:
  a) ratio between the effective centre-of-mass energy and the 
     centre-of-mass energy;
  b) thrust;
  c) $-\ln(y_{4\to3})$;
  d) $\chi^2/n.d.f.$ of the kinematic fit imposing energy-momentum conservation.
  The $\XW\XW$, $\Xq\bar\Xq(\gamma)$ and ``others'' labels represent the 
  background contribution from charged-current four-fermion final 
  states generated with WPHACT~\cite{wphact97,wphact03a,wphact03b}, two-fermion final states 
  generated with KK2F~\cite{kk2f} and all the other processes 
  mentioned in Section~\ref{sec:data}, respectively.
  The signal distributions correspond to scenario $SVT$ (see Table~\ref{tab:scenarios}) and their normalisations are arbitrary, but the same in all plots.
  The arrows show the applied cuts.}
  \label{fig:pre.hadr}
\end{figure}

\begin{figure}[p]
  \begin{center}
    \huge DELPHI
  \end{center}

  \includegraphics[width=.48\textwidth]{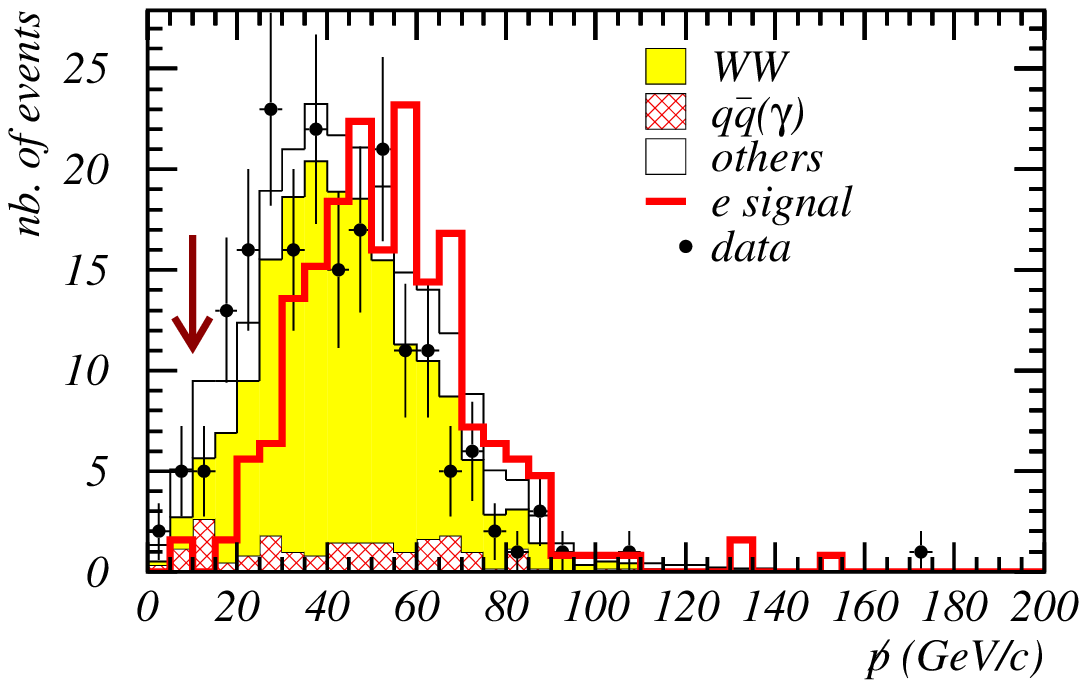}
  \hfill
  \includegraphics[width=.48\textwidth]{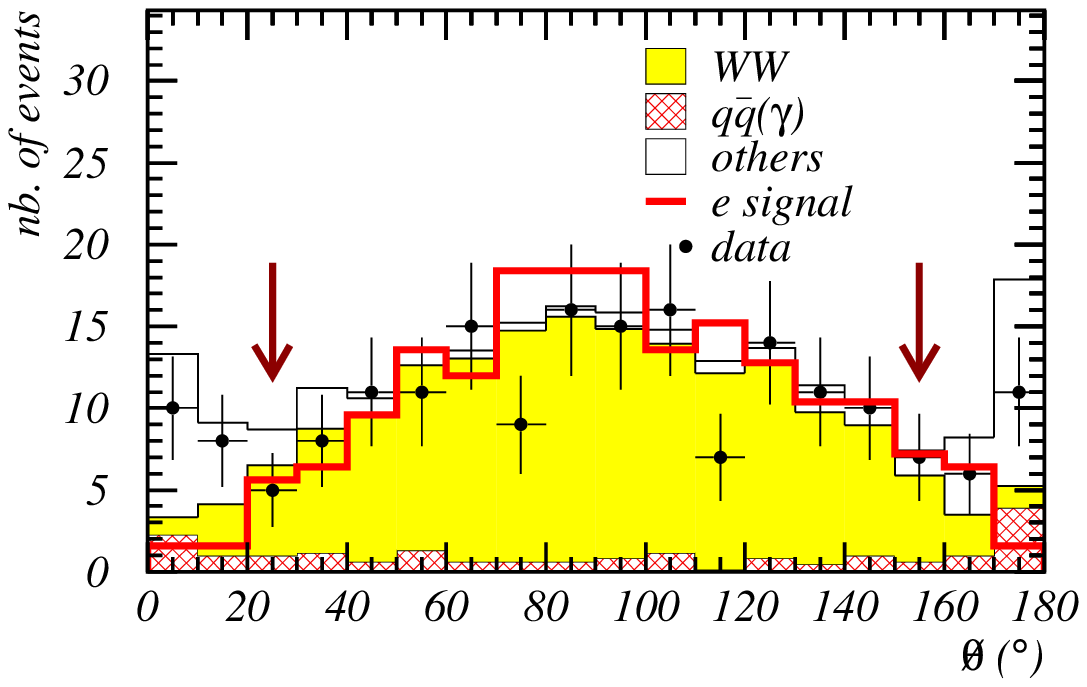}

  \vspace*{-1em}

  \hspace{.24\textwidth}a)\hfill b)\hspace{.24\textwidth}\vspace{1em}

  \includegraphics[width=.48\textwidth]{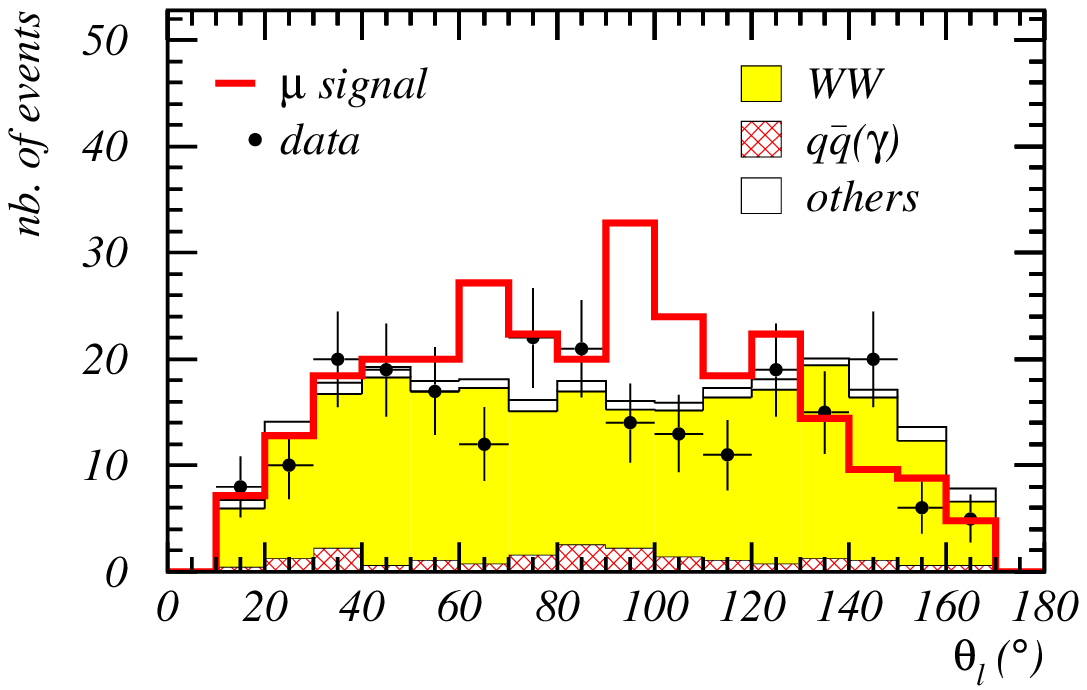}
  \hfill
  \includegraphics[width=.48\textwidth]{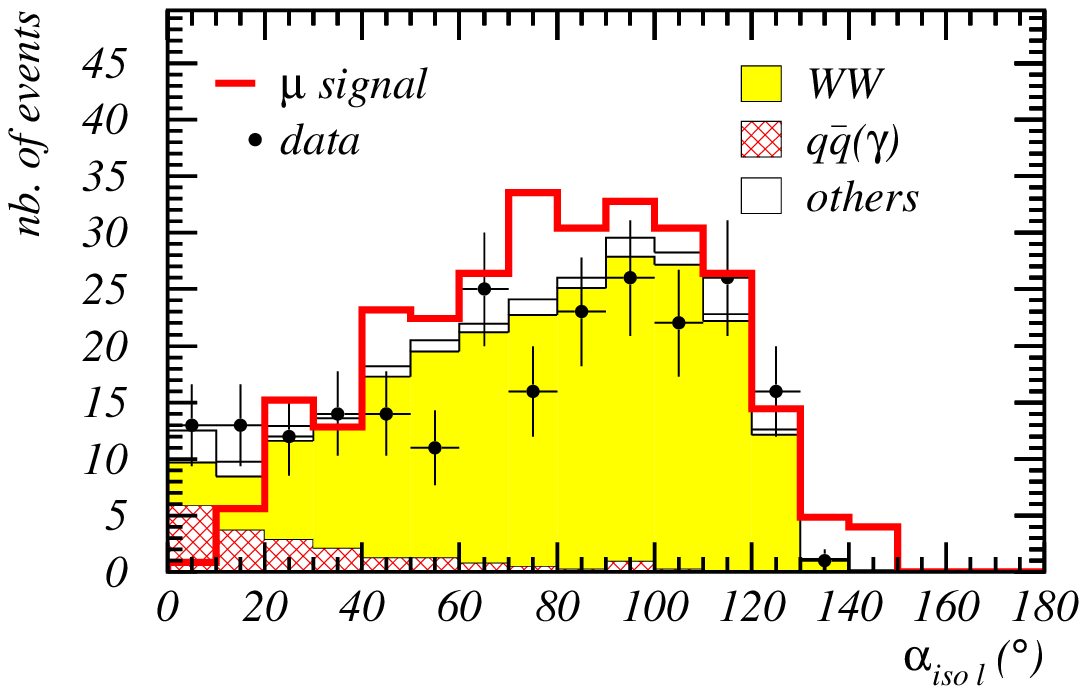}

  \vspace*{-1em}

  \hspace{.24\textwidth}c)\hfill d)\hspace{.24\textwidth}\vspace{1em}

  \includegraphics[width=.48\textwidth]{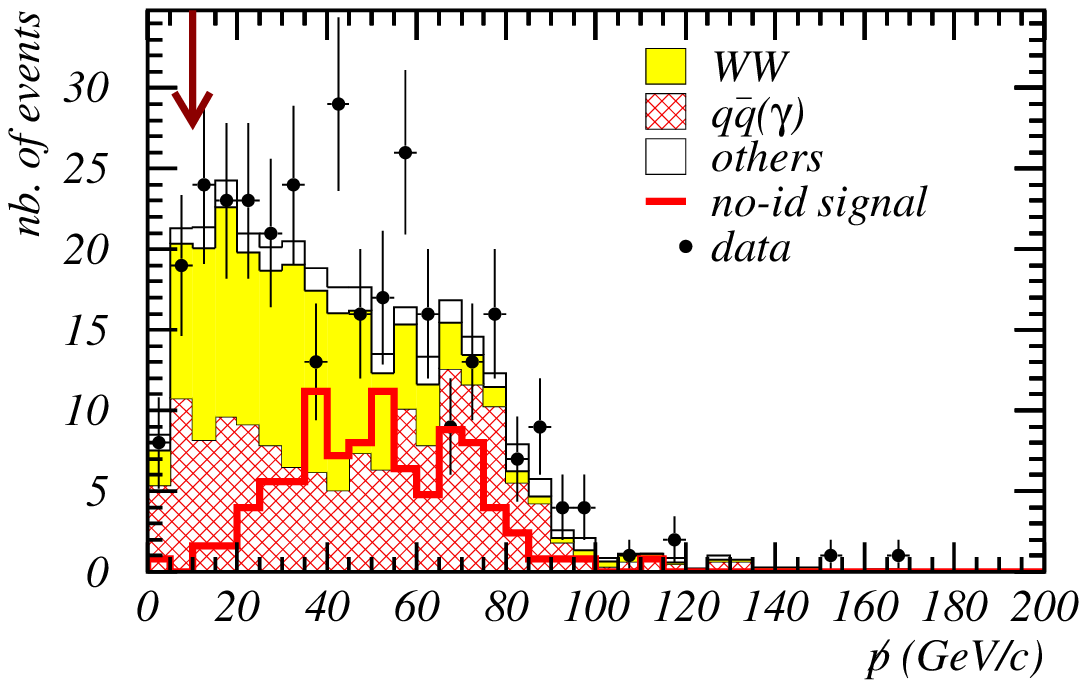}
  \hfill
  \includegraphics[width=.48\textwidth]{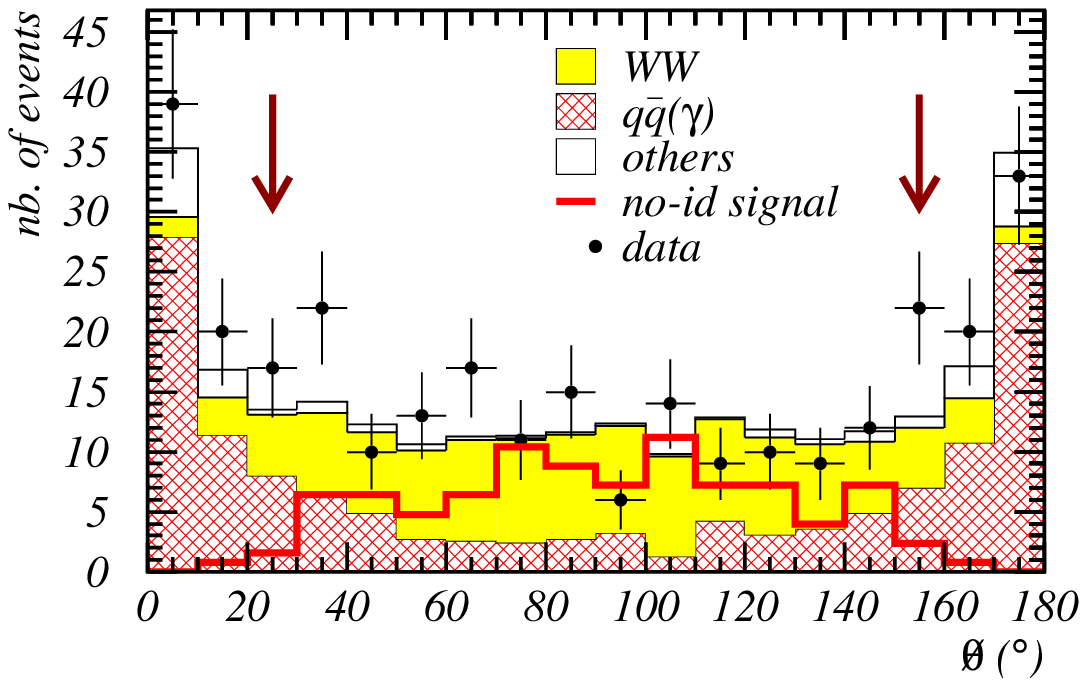}

  \vspace*{-1em}

  \hspace{.24\textwidth}e)\hfill f)\hspace{.24\textwidth}\vspace{1em}
  
  \caption{Distributions of variables relevant for the sequential 
  selection of the semi-leptonic topology after the common preselection 
  are shown at $\langle\sqrt{s}\rangle=206.6$~GeV.
  $e$ sample:
  a) missing momentum;
  b) polar angle of the missing momentum (after applying the cut on the 
  missing momentum distribution);
  $\mu$ sample:
  c) lepton polar angle;
  d) lepton isolation angle;
  \emph{no-id} sample:
  e) missing momentum;
  f) polar angle of the missing momentum (after applying the cut on the 
  missing momentum distribution).
  The $\XW\XW$, $\Xq\bar\Xq(\gamma)$ and ``others'' labels have the same 
  meaning as in Fig.~\ref{fig:pre.hadr}.
  The signal distributions correspond to scenario $SVT$ (see Table~\ref{tab:scenarios}) and their normalisations are arbitrary, but the same in all plots.
  The arrows show the applied cuts.}
  \label{fig:pre.slep}
\end{figure}

\begin{figure}[p]
  \begin{center}
    \huge DELPHI
  \end{center}

  \includegraphics[width=.48\textwidth]{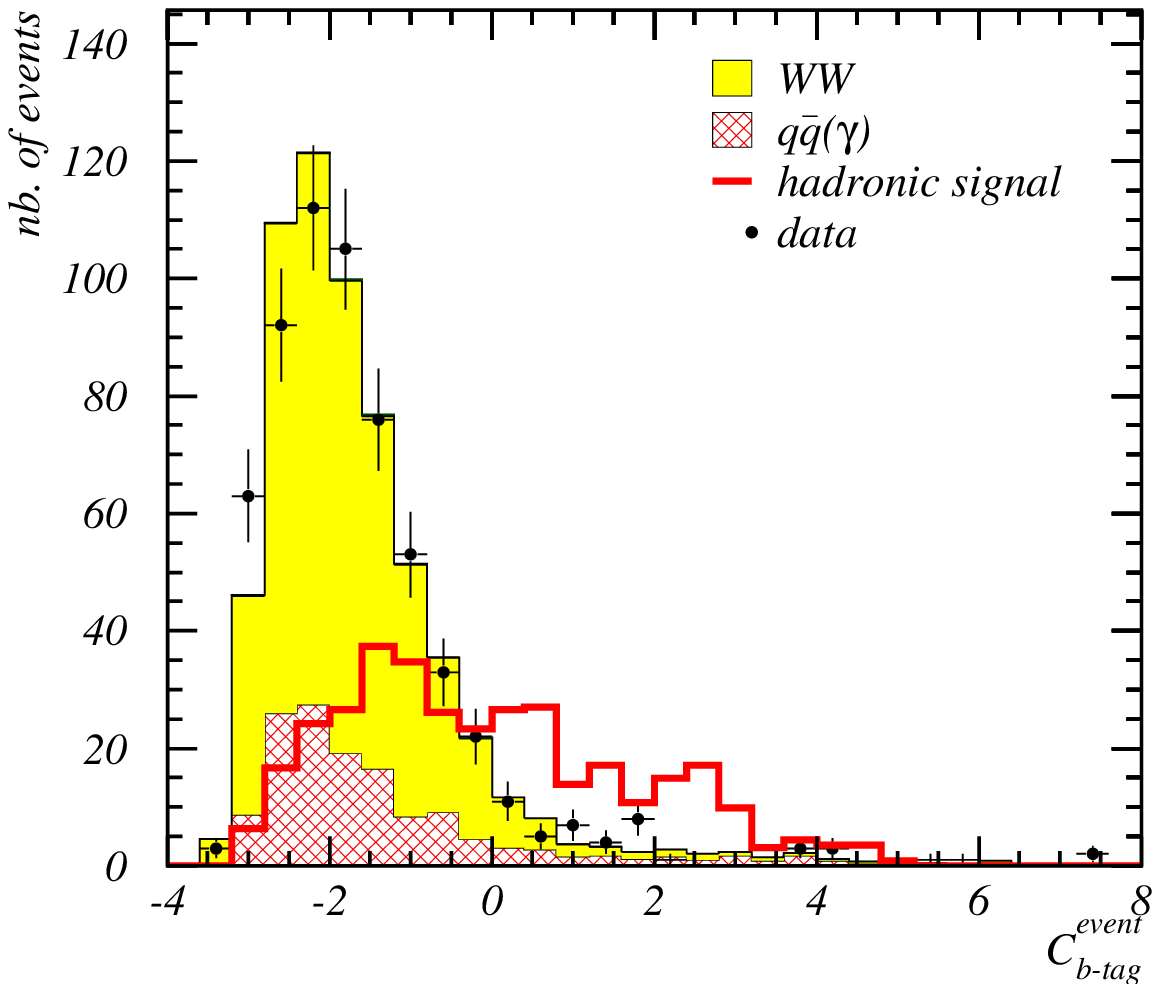}
  \hfill
  \includegraphics[width=.48\textwidth]{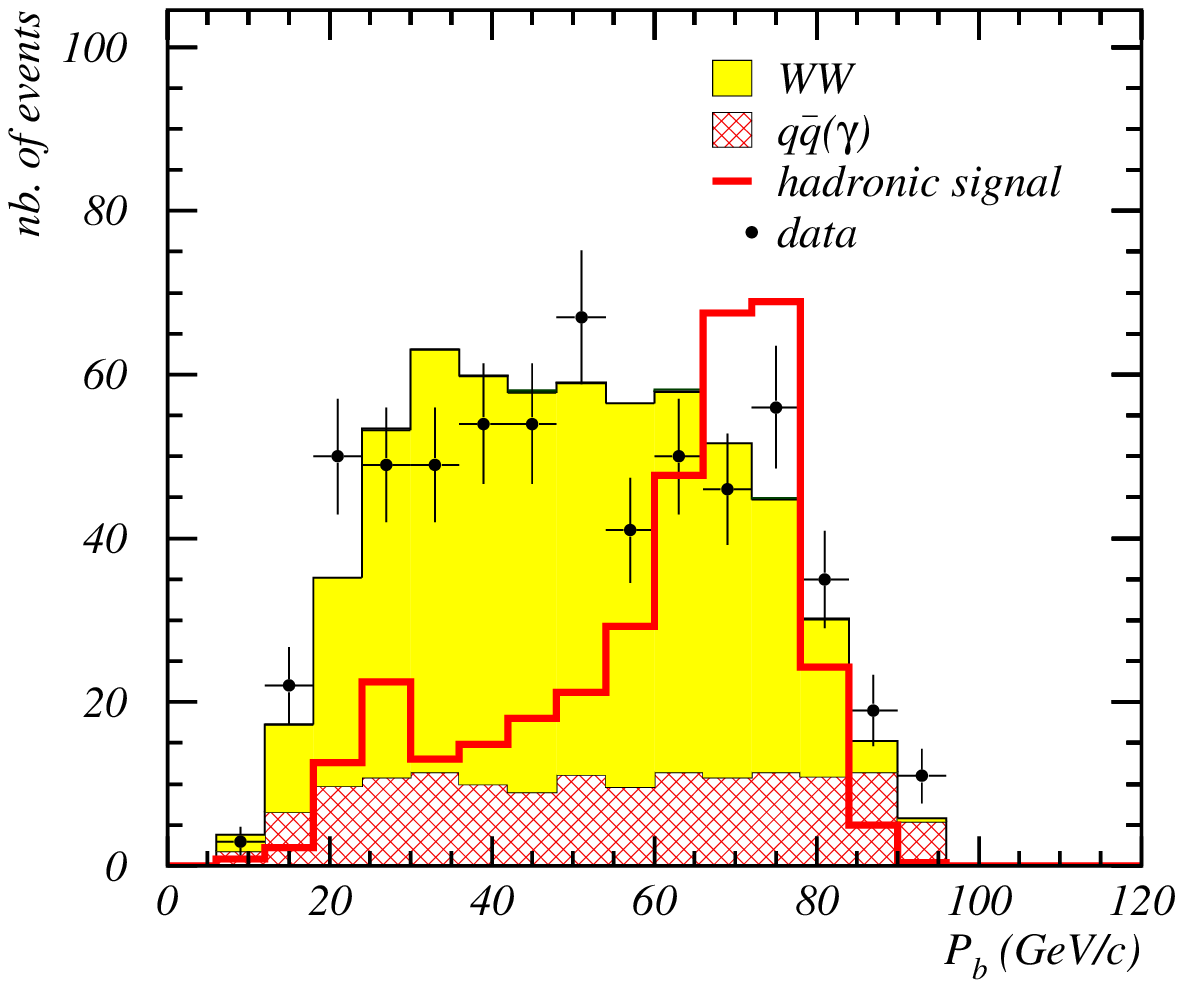}

  \vspace*{-1.5em}

  \hspace{.24\textwidth}a)\hfill b)\hspace{.24\textwidth}\vspace{1.5em}

  \includegraphics[width=.48\textwidth]{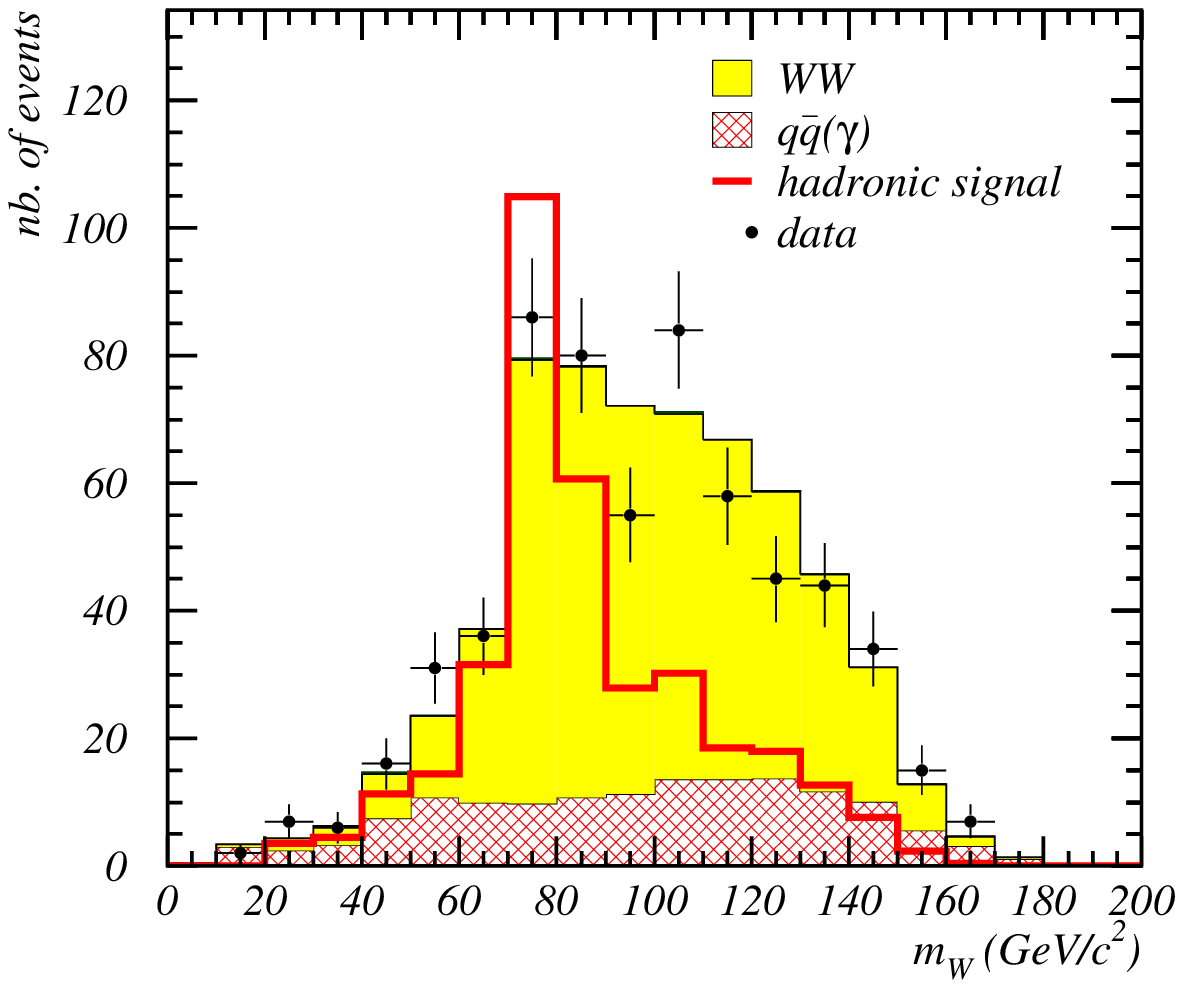}
  \hfill
  \includegraphics[width=.48\textwidth]{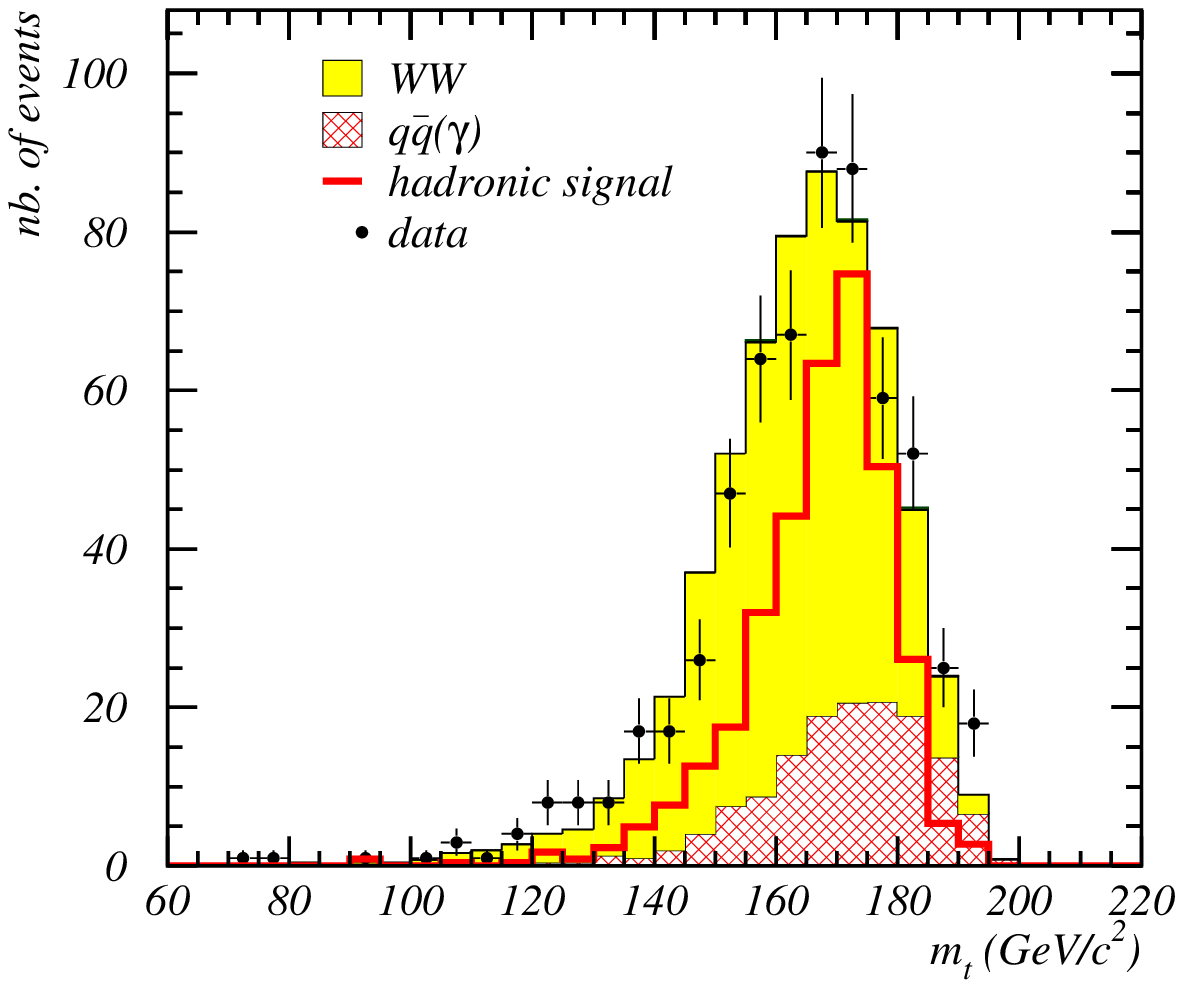}

  \vspace*{-1.5em}

  \hspace{.24\textwidth}c)\hfill d)\hspace{.24\textwidth}\vspace{1.5em}

  \caption{Distributions of variables relevant for the 
  hadronic topology after 
  the sequential selection at $\langle\sqrt{s}\rangle=206.6$~GeV: 
  a) $\Xb$-tag of the event;
  b) $\Xb$ jet momentum;
  c) reconstructed $\XW$ boson mass;
  d) reconstructed $\Xt$ quark mass.
  The a), b) and c) distributions were used as PDF to construct the 
  discriminant variable for the hadronic topology.
  The signal distributions correspond to scenario $SVT$ (see Table~\ref{tab:scenarios}) and their normalisations are arbitrary, but the same in all plots.}
  \label{fig:pdf.hadr}
\end{figure}

\begin{figure}[p]
  \begin{center}
    \huge DELPHI
  \end{center}

  \includegraphics[width=.48\textwidth]{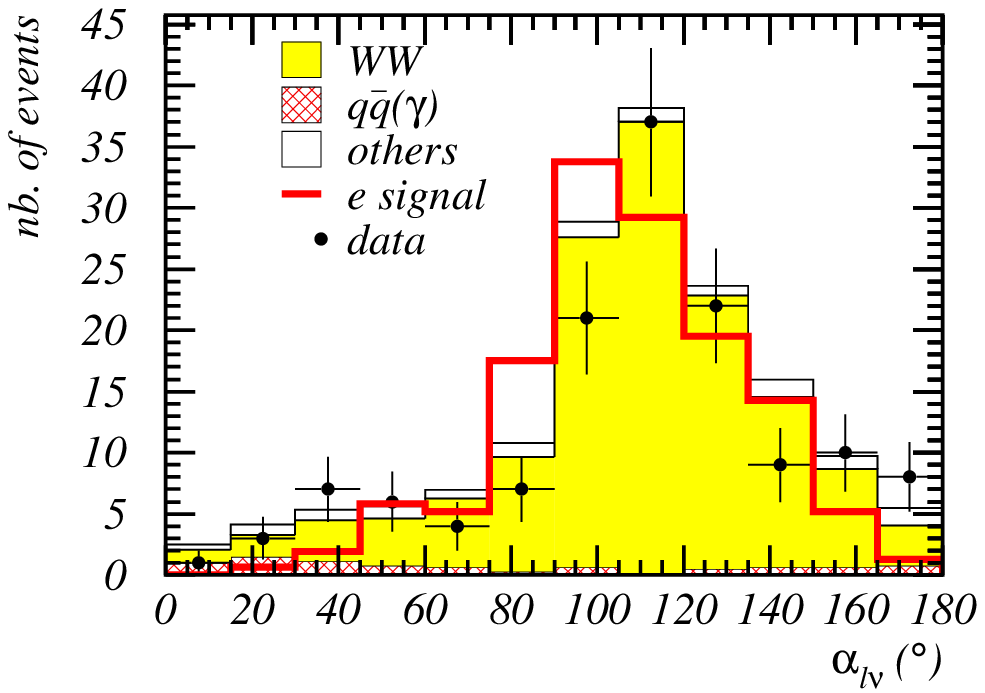}
  \hfill
  \includegraphics[width=.48\textwidth]{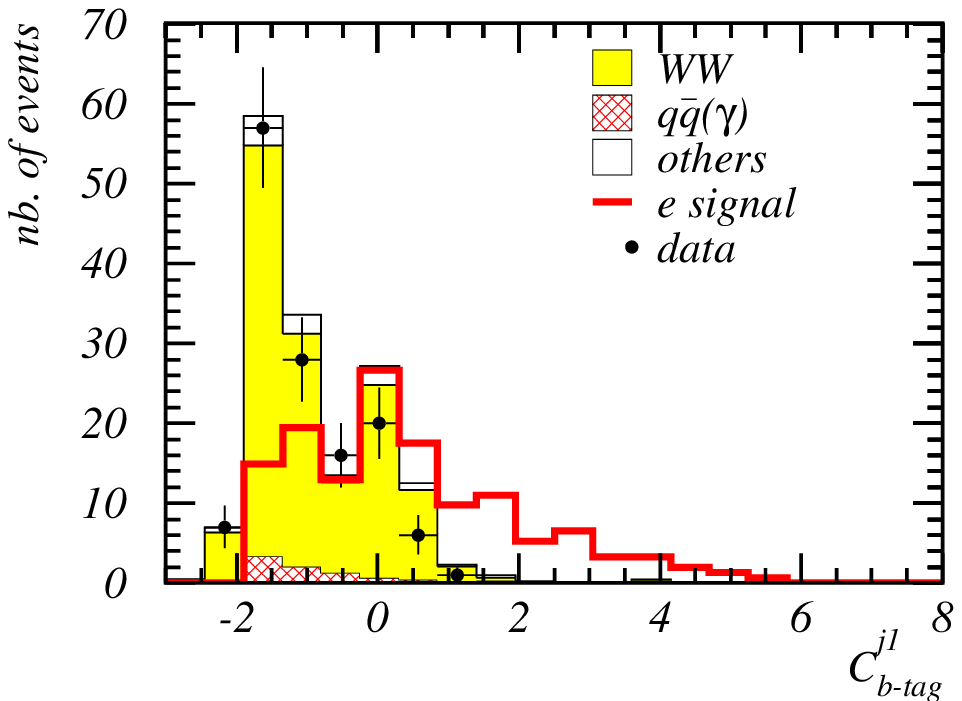}

  \vspace*{-1.5em}

  \hspace{.24\textwidth}a)\hfill b)\hspace{.24\textwidth}\vspace{1.5em}

  \includegraphics[width=.48\textwidth]{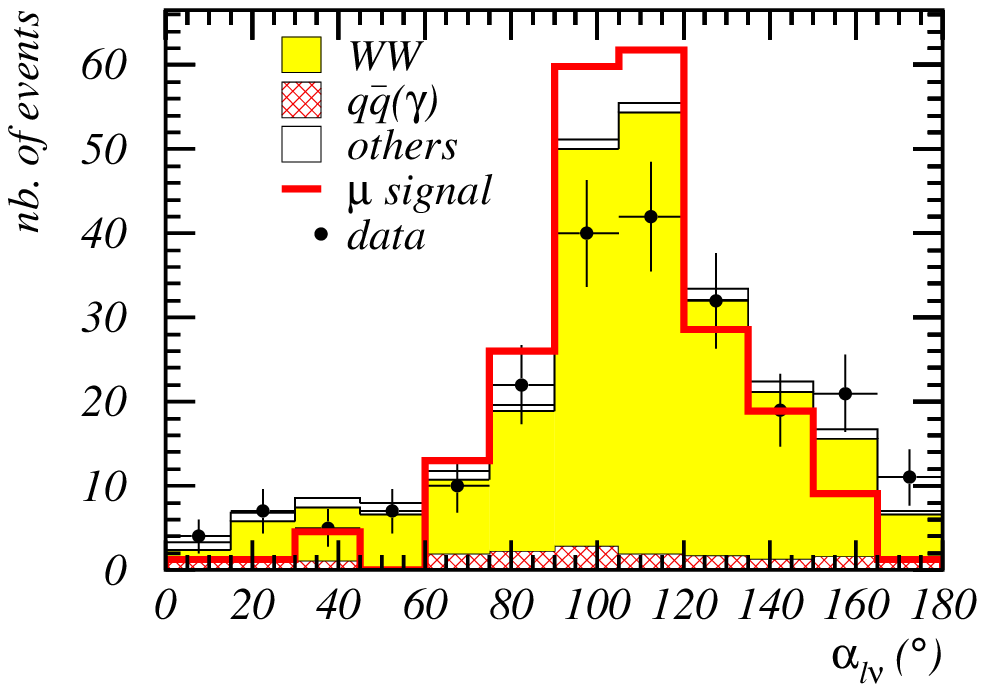}
  \hfill
  \includegraphics[width=.48\textwidth]{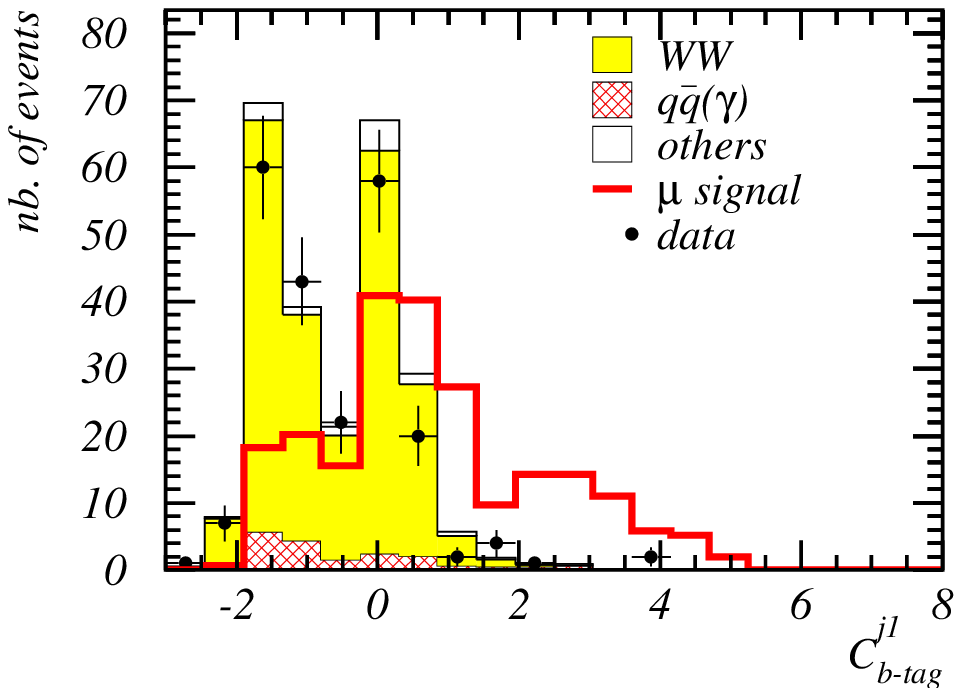}

  \vspace*{-1.5em}

  \hspace{.24\textwidth}c)\hfill d)\hspace{.24\textwidth}\vspace{1.5em}

  \includegraphics[width=.48\textwidth]{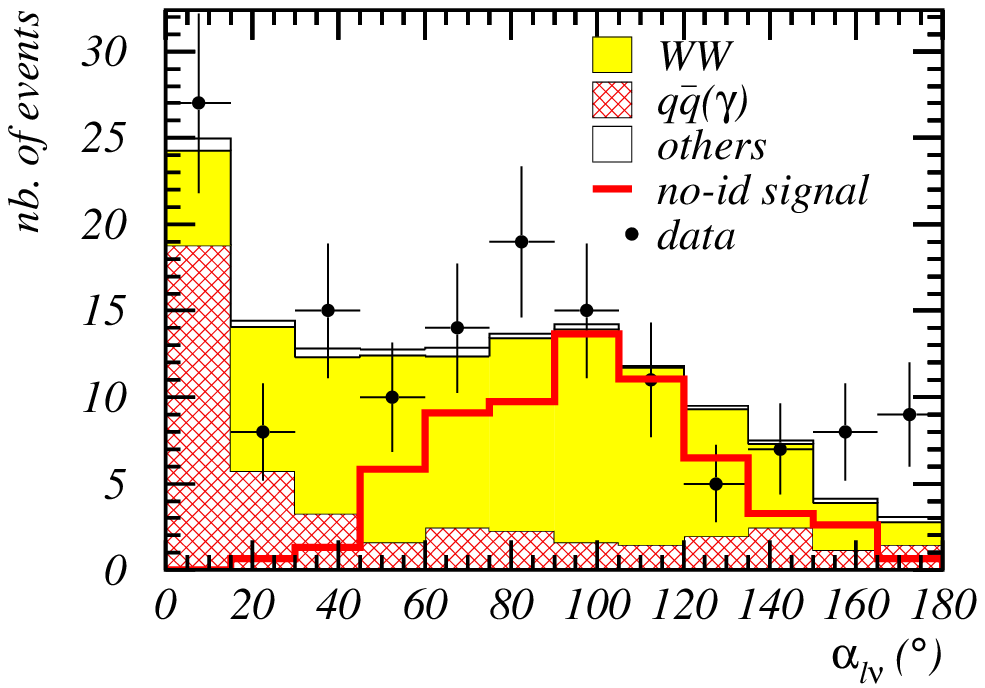}
  \hfill
  \includegraphics[width=.48\textwidth]{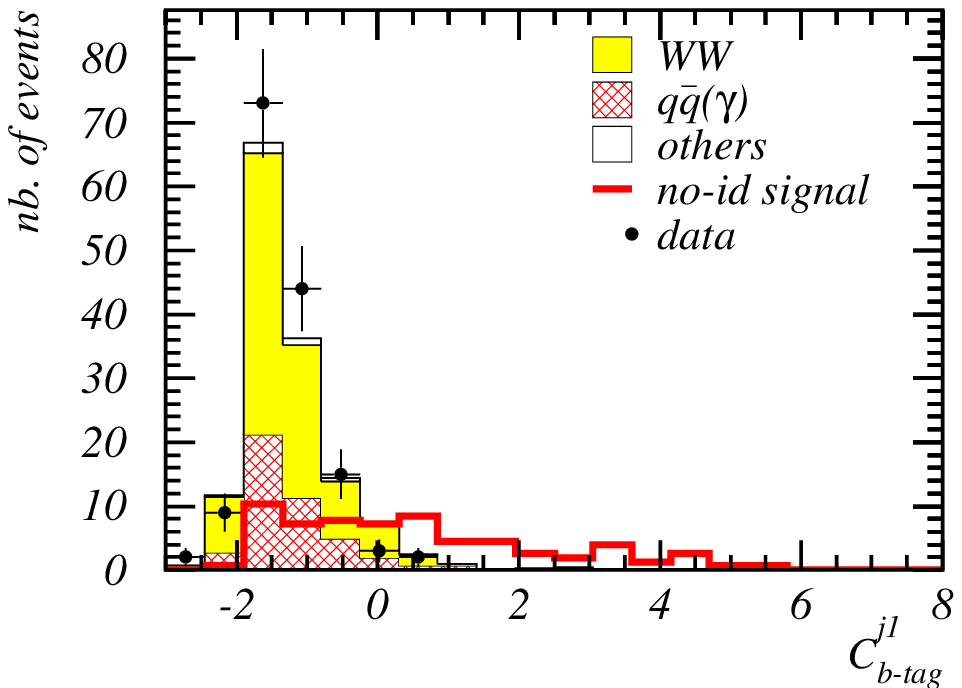}

  \vspace*{-1.5em}

  \hspace{.24\textwidth}e)\hfill f)\hspace{.24\textwidth}\vspace{1.5em}
  
  \caption{Distributions of variables relevant for the semi-leptonic 
  topology after the sequential selection at
  $\langle\sqrt{s}\rangle=206.6$~GeV 
  In the left column: angle between the lepton and the neutrino;
  in the right column: $\Xb$-tag of most energetic jet;
  (a,b) $\Xe$ sample;
  (c,d) $\mu$ sample;
  (e,f) \emph{no-id} sample.
  The signal distributions correspond to scenario $SVT$ (see Table~\ref{tab:scenarios}) and their normalisations are arbitrary, but the same in all plots.}
  \label{fig:pdf.slep}
\end{figure}

\begin{figure}[p]
  \begin{center}
    \huge DELPHI
  \end{center}

  \includegraphics[width=.48\textwidth]{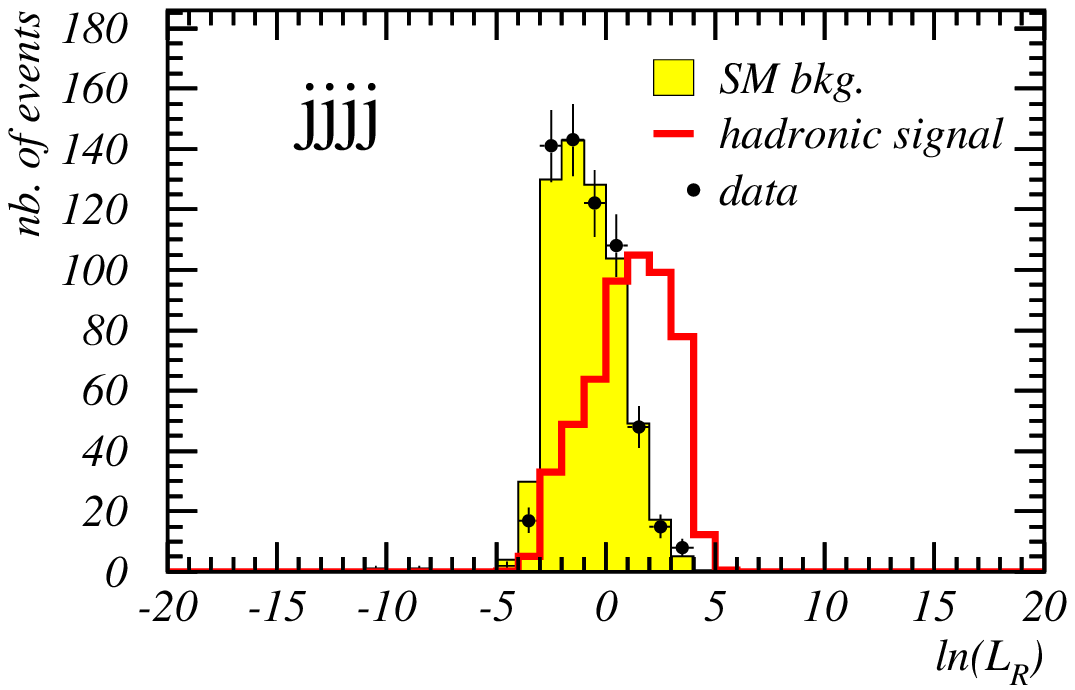}
  \hfill
  \includegraphics[width=.48\textwidth]{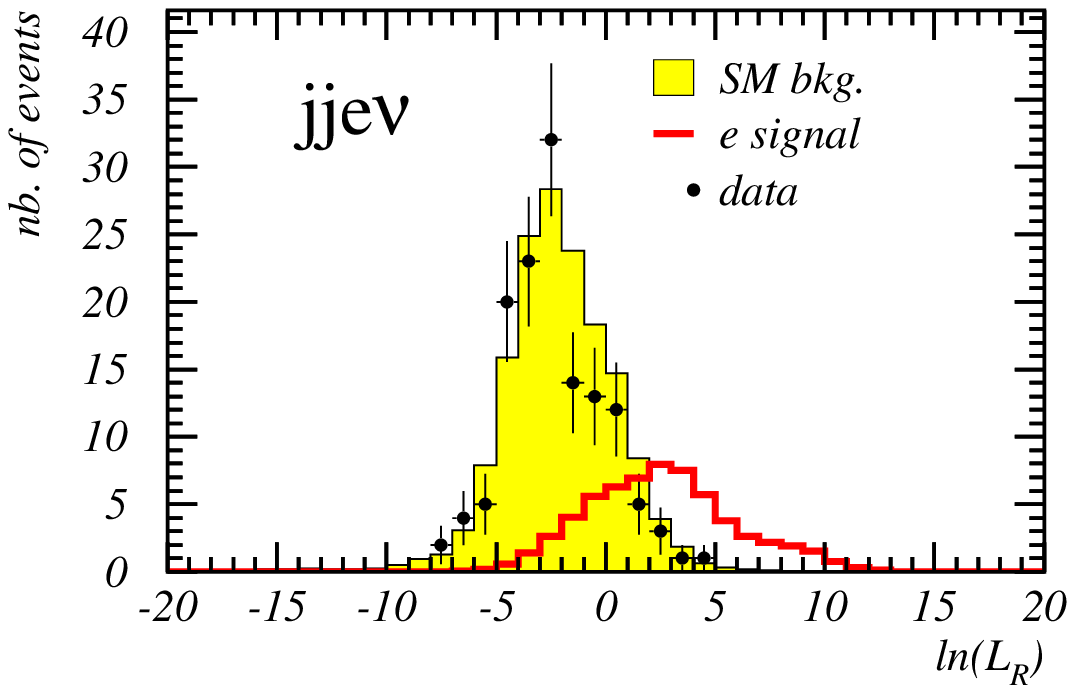}

  \vspace*{-1.25em}

  \hspace{.24\textwidth}a)\hfill b)\hspace{.24\textwidth}\vspace{1.25em}



  \includegraphics[width=.48\textwidth]{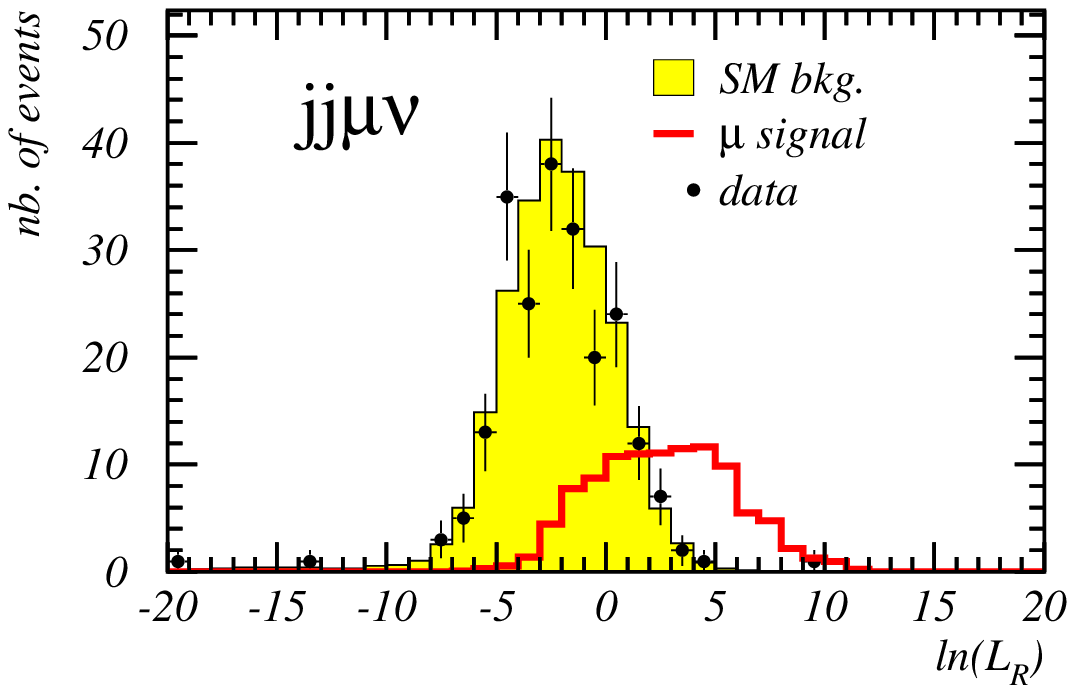}
  \hfill
  \includegraphics[width=.48\textwidth]{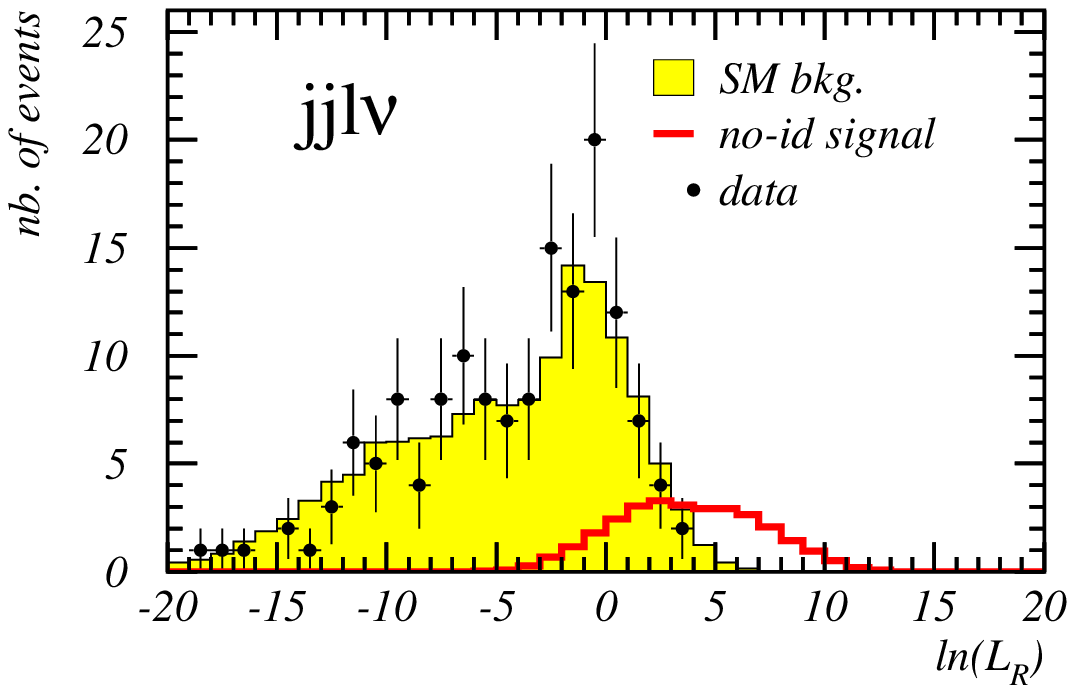}

  \vspace*{-1.25em}

  \hspace{.24\textwidth}c)\hfill d)\hspace{.24\textwidth}\vspace{1.25em}
  


  \caption{Distributions of the discriminant variable 
  $\ln{\mathcal{L}}_R$ for data, expected background and signal
  after the sequential selection at
  $\langle\sqrt{s}\rangle=206.6$~GeV:
  a) hadronic topology;
  semi-leptonic topology:
  b) $\Xe$ sample;
  c) $\mu$ sample;
  d) \emph{no-id} sample.
  These distributions correspond to scenario $SVT$ (see Table~\ref{tab:scenarios}).
  The signal normalisation is arbitrary, but the same in all plots.}
\label{fig:discriminant}
\end{figure}

The final state of the hadronic topology ($\Xt\bar \Xc\to \Xb\bar
\Xc\Xq\bar {\Xq^\prime}$) is characterized by the presence of four jets,
two of them from the $\XW$ hadronic decay, and no missing energy. The
distributions of relevant variables after the common preselection are
shown\footnote{For illustration purposes all the simulated signal distributions in
Figs.~\ref{fig:pre.hadr}--\ref{fig:pdf.slep} and all the plotted
distributions in Fig.~\ref{fig:discriminant} correspond to scenario $SVT$ 
(see Table~\ref{tab:scenarios}).} in
Fig.~\ref{fig:pre.hadr}. Due to the high multiplicity expected in this
topology, the required minimum number of good charged-particles tracks was
raised to 25. The events were required to have an effective centre-of-mass
energy~\cite{sprime} $\sqrt{s^\prime}\ge 0.7\sqrt s$ and a thrust below
0.9.  Events were clustered into four jets and it was required that
$-\ln(y_{4\to 3})\le6.5$, where $y_{n\to n-1}$ is the Durham resolution
variable in the transition from $n$ to $n-1$ jets. Assuming a four-jets
final state, a kinematic fit requiring energy-momentum conservation was
performed. Events with $\chi^2/n.d.f.$ lower than 10 were accepted.

In Table \ref{tab:sequential} the number of selected data events, the 
expected SM background and the signal efficiencies at the end of the 
sequential selection are shown for the different centre-of-mass energies.

\begin{table}[p]
  \begin{center}\begin{tabular*}{\textwidth}{@{\extracolsep{\fill}}lrrrrrrrr}
    \hline
    \hline
    $\langle\sqrt{s}\rangle$ (GeV) & 188.6 & 191.6 & 195.5 & 199.5 & 201.6 & 204.8 & 206.6 & 206.3$^*$ \\
    \hline
    \multicolumn{9}{l}{Hadronic topology:}\\
    data                    &    1165\Xo &    211\Xo &    613\Xo &    637\Xo &    306\Xo &    599\Xo &    606\Xo &    400\Xo \\
    background              &    1216.1  &    197.0  &    589.5  &    637.7  &    299.6  &    610.6  &    612.7  &    444.1  \\
                            & $\pm$14.4  & $\pm$2.3  & $\pm$6.6  & $\pm$7.0  & $\pm$3.3  & $\pm$6.6  & $\pm$6.5  & $\pm$4.8  \\
    $\varepsilon$ min. (\%) &      46.5  &     42.8  &     42.8  &     50.9  &     50.9  &     51.5  &     51.5  &     50.5  \\
    $\varepsilon$ max. (\%) &      48.2  &     48.9  &     48.9  &     54.0  &     54.0  &     55.6  &     55.6  &     54.5  \\
    \hline
    \multicolumn{9}{l}{Semi-leptonic topology -- $\Xe$ sample:}\\
    data                    &     259\Xo &     37\Xo &    140\Xo &    151\Xo &     80\Xo &    166\Xo &    137\Xo &    106\Xo \\
    background              &     290.8  &     46.0  &    142.8  &    157.1  &     75.9  &    155.2  &    158.2  &    109.6  \\
                            &  $\pm$5.2  & $\pm$0.8  & $\pm$2.5  & $\pm$2.8  & $\pm$1.3  & $\pm$2.7  & $\pm$2.8  & $\pm$2.0  \\
    $\varepsilon$ min. (\%) &       6.5  &      6.1  &      6.1  &      6.4  &      6.4  &      6.5  &      6.5  &      6.2  \\
    $\varepsilon$ max. (\%) &       7.6  &      7.3  &      7.3  &      7.2  &      7.2  &      7.1  &      7.6  &      7.1  \\  
    \hline
    \multicolumn{9}{l}{Semi-leptonic topology -- $\mu$ sample:}\\
    data                    &     423\Xo &     75\Xo &    226\Xo &    259\Xo &    111\Xo &    240\Xo &    220\Xo &    169\Xo \\
    background              &     432.9  &     75.4  &    225.6  &    246.7  &    118.4  &    232.8  &    244.3  &    169.9  \\
                            &  $\pm$6.5  & $\pm$1.1  & $\pm$3.3  & $\pm$3.6  & $\pm$1.7  & $\pm$3.3  & $\pm$3.5  & $\pm$2.5  \\
    $\varepsilon$ min. (\%) &      10.6  &     10.5  &     10.5  &     10.3  &     10.3  &     10.7  &     10.5  &      9.9  \\
    $\varepsilon$ max. (\%) &      11.5  &     11.6  &     11.6  &     11.4  &     11.4  &     11.1  &     11.5  &     10.8  \\
    \hline
    \multicolumn{9}{l}{Semi-leptonic topology -- \emph{no-id} sample:}\\
    data                    &     308\Xo &     49\Xo &    140\Xo &    135\Xo &     67\Xo &    145\Xo &    148\Xo &     92\Xo \\
    background              &     286.2  &     45.4  &    133.9  &    146.8  &     72.0  &    141.1  &    141.7  &    104.5  \\
                            &  $\pm$7.5  & $\pm$1.2  & $\pm$3.3  & $\pm$3.6  & $\pm$1.7  & $\pm$3.3  & $\pm$3.4  & $\pm$2.5  \\
    $\varepsilon$ min. (\%) &       2.7  &      2.6  &      2.6  &      2.8  &      2.8  &      2.9  &      3.3  &      2.7  \\
    $\varepsilon$ max. (\%) &       3.5  &      3.4  &      3.4  &      3.3  &      3.3  &      3.3  &      3.6  &      3.4  \\
    \hline
    \multicolumn{9}{l}{Total:}\\
    data                    &    2155\Xo &    372\Xo &   1119\Xo &   1182\Xo &    564\Xo &   1150\Xo &   1111\Xo &    767\Xo \\
    background              &    2226.0  &    363.8  &   1091.8  &   1188.3  &    565.9  &   1139.7  &   1156.9  &    828.1  \\
                            & $\pm$18.2  & $\pm$2.9  & $\pm$8.5  & $\pm$9.1  & $\pm$4.3  & $\pm$8.5  & $\pm$8.6  & $\pm$6.3  \\
    $\varepsilon$ min. (\%) &      67.5  &     62.3  &     62.3  &     71.3  &     71.3  &     72.7  &     72.6  &     69.8  \\
    $\varepsilon$ max. (\%) &      69.3  &     69.8  &     69.8  &     75.0  &     75.0  &     76.7  &     77.5  &     73.6  \\
    \hline
    \hline
  \end{tabular*}\end{center}
  \caption{Number of selected data events, expected SM background and 
  signal efficiencies, $\epsilon$, (in percent and convoluted with 
  the branching ratio of the $\XW$ boson) after the sequential 
  selection. The expected background numbers are shown with their 
  statistical errors. The efficiencies have been computed for the different 
  coupling scenarios according to Table~\ref{tab:scenarios} and the 
  extreme values are shown. The statistical errors on the efficiency 
  are smaller than 1.3\% and 0.6\% for the hadronic and semi-leptonic 
  topologies, respectively.
  \label{tab:sequential}}
\end{table}


\subsubsection*{Semi-leptonic topology}

The final state for the semi-leptonic topology ($\Xt\bar \Xc\to \Xb\bar
\Xc\ell^+\nu_\ell$) is characterised by the presence of two jets, one
isolated and energetic lepton and missing energy. The $\Xb$ jet is
expected to be energetic, while the $\Xc$ jet has low momentum. Events
with at least one isolated charged lepton and at least six good
charged-particles tracks with TPC information were accepted. The
particles of the events, excluding the isolated leptons, were clustered
into two jets using the Durham algorithm and the events were divided
into three mutually exclusive samples according to the identified flavour of the
most energetic lepton: events with a \emph{tight} electron (``$\Xe$
sample''), events with a \emph{standard} or \emph{tight} muon (``$\mu$
sample'') and events in which no unambiguous lepton was present
(``\emph{no-id} sample'')\footnote{Notice that, according to these
definitions, the $\Xe$ and $\mu$ samples also contain the tauonic events
in which the $\tau$ decayed, respectively, into a $\Xe$ (if classified
as \emph{tight}) or a $\mu$ (if classified as \emph{standard} or
\emph{tight}).}.

In the $\Xe$ and \emph{no-id} samples, photons converting in the tracking 
system were vetoed by requiring that the lepton had left a signal in at least 
two layers of the vertex detector. Contamination from Bhabha and 
$\gamma\gamma$ events was reduced in the $\Xe$ sample by imposing that the 
lepton was above $25^\circ$ in polar angle and that the isolation angle, 
defined as the angle between the lepton and the nearest jet, isolated photon 
or other isolated charged leptons, was greater than $10^\circ$. For the 
\emph{no-id} sample, the contribution from these backgrounds was reduced 
by keeping only events with exactly one isolated lepton with a polar angle 
greater than $25^\circ$, momentum above $0.075\sqrt s$ and an isolation angle 
of at least $20^\circ$. The distributions of relevant variables after these 
cuts are shown in Fig.~\ref{fig:pre.slep}. The $\Xq\bar \Xq$ background 
contamination, in the $\Xe$ and \emph{no-id} samples, was further reduced by 
requiring a missing momentum above 10~GeV/$c$ pointing above $25^\circ$ in 
polar angle.

Assuming a $\Xj\Xj\ell\nu$ final state and assigning the missing
momentum to the undetected neutrino, a kinematic fit requiring
energy-momentum conservation was applied in all three samples. Events
with $\chi^2/n.d.f.$ lower than 10 were accepted.

In Table 3 the number of selected data events, the expected SM
background and the signal efficiencies are shown at the end of the
sequential selection.


\subsection{Discriminant selection\label{sec:discriminant}}

After the sequential analysis, the main background consisted of
$\XW^+\XW^-$ events, which are similar to the signal and have the same
final state topology. A separation is possible, based on the different
kinematic properties and on jet-flavour tagging techniques.
Furthermore, the $\XW$ and $\Xt$ quark
mass constraints can be used to separate signal and background.


\subsubsection*{Hadronic topology}

In order to identify the $\Xb$ and $\Xc$ jets and determine the kinematic
properties of the $\Xt$ quark and of the $\XW$ boson, several possible jet
assignment schemes were studied:
\begin{enumerate}
  \item the most energetic jet is assigned to the $\Xb$ quark and the 
    least energetic one to the $\Xc$ quark;
  \item the most energetic jet is assigned to the $\Xb$ quark and the
    jets that minimise $|m_{\Xj\Xj}-m_\XW|$, where $m_{\Xj\Xj}$ is the 
    invariant mass of two of the three remaining jets and
    $m_\XW=80.4$~GeV$/c^2$, are assigned to the $\XW$ boson;
  \item the jet with the highest $\Xb$-tag value is assigned to the $\Xb$ quark 
    and the least energetic one of the remaining three to the $\Xc$ quark; 
  \item the jet with the highest $\Xb$-tag value is assigned to the $\Xb$ quark 
    and, from the three remaining, the jets that minimise
    $|m_{\Xj\Xj}-m_\XW|$ are assigned to the $\XW$ boson.
\end{enumerate}

The correct assignment of jets to quarks was studied with simulated
signal samples at $\sqrt{s}=189$~GeV and $\sqrt{s}=206$~GeV. Correct
assignment was defined based on the angle $\alpha_{\Xq\Xj}$ between the
quark and jet direction, requiring $\cos\alpha_{\Xq\Xj}\geq0.9$. The
results are presented in Table~\ref{tab:jets.attr.hadr}: higher
efficiencies for the $\Xb$ quark assignment are obtained with schemes
3 and 4. Scheme 3 was adopted since it also yields the best discrimination
between signal and background.

Signal and background-like probabilities were assigned to each event
based on Probability Density Functions~(PDF) constructed with the
following variables:
\begin{itemize}
  \item the event $\Xb$-tag value, 
  $C_{\Xb\hbox{-}\mathrm{tag}}^{\mathrm{event}}$;
  \item the $\Xb$ momentum, $P_\Xb$;
  \item the $\XW$ reconstructed mass, $m_\XW$.
\end{itemize}
The distributions of these variables are shown in
Fig.~\ref{fig:pdf.hadr} for data, expected background and signal at
$\langle\sqrt{s}\rangle=206.6$~GeV. For each of these variables, the
corresponding PDF for the signal ($P_S^i$) and background ($P_B^i$) were
estimated. For each event, a signal likelihood ($P_S=\prod_i P_S^i$) and
background likelihood ($P_B=\prod_i P_B^i$) were computed assuming no
correlations. The discriminant variable was defined as
$\ln{\mathcal{L}}_R=\ln(P_S/P_B)$.


\begin{table}[tb]
  \begin{center}\begin{tabular*}{\textwidth}{@{\extracolsep{\fill}}lrrrr}
    \hline
    \hline
    Scheme:              &  1   &  2   &  3   &  4   \\
    \hline
    $\sqrt{s}=189$~GeV: \\
    $\varepsilon_b$ (\%): & 52.4 & 52.4 & 72.5 & 72.5 \\
    $\varepsilon_c$ (\%): & 45.5 & 43.4 & 41.9 & 40.7 \\
    \hline
    $\sqrt{s}=206$~GeV: \\
    $\varepsilon_b$ (\%): & 53.3 & 53.3 & 68.0 & 68.0 \\
    $\varepsilon_c$ (\%): & 51.2 & 51.0 & 47.1 & 44.8 \\
    \hline
    \hline
  \end{tabular*}\end{center}
  \caption{Fraction of the correct assignments of jets to quarks for 
  simulated signal events of the hadronic topology at $\sqrt{s}=$~189~GeV 
  and $\sqrt{s}=$~206~GeV, using the four jet assignment schemes 
  explained in the text.}
  \label{tab:jets.attr.hadr}
\end{table}


\subsubsection*{Semi-leptonic topology}

A discriminant variable was constructed using signal ($P_S^i$)
and background ($P_B^i$) PDFs estimated from the following variables:
\begin{itemize}
  \item the angle between the two jets, $\alpha_{j_1j_2}$;
  \item the angle between the lepton and the neutrino, $\alpha_{\ell\nu}$;
  \item the reconstructed mass of the two jets, $m_{j_1j_2}$;
  \item the reconstructed mass of the jet with the highest $\Xb$-tag, 
        the lepton and the neutrino, $m_{\Xb\ell\nu}$;
  \item the reconstructed $\XW$ mass, $m_{\ell\nu}$;
  \item the ratio of the jet momenta, $P_{j_2}/P_{j_1}$;
  \item the $\Xb$-tag of the most energetic jet,
        $C_{\Xb\hbox{-}\mathrm{tag}}^{j_1}$;
  \item the product of the lepton charge and the cosine of the lepton polar 
        angle, $Q_\ell\times\cos\theta_\ell$;
  \item $-\ln y_{2\to1}$, where $y_{2\to 1}$ is the Durham 
        resolution variable in the transition from two to one jet.
\end{itemize}

Distributions of some of these variables are shown in
Fig.~\ref{fig:pdf.slep} for $\langle\sqrt{s}\rangle=206.6$~GeV.



\section{Results\label{sec:results}}

The discriminant variables obtained in the different search channels are
shown in Fig.~\ref{fig:discriminant}, for
$\langle\sqrt{s}\rangle=206.6$~GeV. As no signal was found in any of the
analysis channels, limits at 95\% confidence level (CL) on the energy
scale $\Lambda$ were derived for each of the scenarios in
Table~\ref{tab:scenarios}. The limits were obtained using the modified
frequentist likelihood ratio method~\cite{alexread}, taking into account
the observed and expected event counts, the signal efficiencies and the
shapes of the discriminant variables in data, background and signal. The
expected limit was computed as the median limit for experiments without
signal contributions. The $\pm 1\sigma$ values around the expected
median limit were also computed. In order to avoid non-physical
fluctuations in the distributions of the discriminant variables, due to
the limited statistics of the generated events, a smoothing procedure
was adopted. The limits were evaluated assuming $m_\Xt=175$~GeV$/c^2$,
which allows direct comparison with other published results. The
results, obtained with the contribution of all the systematic
uncertainties described in the next paragraph, are presented in
Table~\ref{tab:limits}. The observed and expected limits are
statistically compatible and the maximum difference is about 1$\sigma$.
The effect of a change in the $\Xt$ quark mass was studied at two
extreme energies (188.6 and 204.8~GeV) and two extreme scenarios ($SVT$
and $a$), considering the values 170 and 180 GeV$/c^2$ for $m_\Xt$. The
estimated relative changes in the limits were about $+10$\% for
$m_\Xt=170$~GeV$/c^2$ and $-14$\% for $m_\Xt=180$~GeV$/c^2$. For
scenarios $S$ and $T$, in which only one coupling is non-zero, limits at
95\% CL on the ratio between the coupling and $\Lambda^2$ can be
directly obtained from Equ.~\ref{eq:sig}:
\begin{displaymath}
  \left({T\over\Lambda^2}\right)_{\mathrm{obs}}\le
    6.90\times10^{-7}\mathrm{GeV}^{-2},
  \quad
  \left({T\over\Lambda^2}\right)_{\mathrm{exp}}\le
    6.83\times10^{-7}\mathrm{GeV}^{-2},
\end{displaymath}
\begin{displaymath}
  \left({S\over\Lambda^2}\right)_{\mathrm{obs}}\le
    2.13\times10^{-6}\mathrm{GeV}^{-2},
  \quad
  \left({S\over\Lambda^2}\right)_{\mathrm{exp}}\le
    2.43\times10^{-6}\mathrm{GeV}^{-2}.
\end{displaymath}
The limit obtained in scenario $a$, involving only the $a^Z_j$
couplings, can be converted into a limit on the anomalous coupling
$\kappa_\XZ$ (see Equ.~\ref{eq:akz})\footnote{Notice that in
Ref.~\cite{aleph00,aleph02,fcnc,l3,opal} the contribution from the processes
$\Xe^+\Xe^-\to \Xt\bar \Xu$ and $\Xe^+\Xe^-\to \bar\Xt\Xu$ are also
considered. This was taken into account in the limits conversion.},
yielding $\kappa_\XZ^{\mathrm{obs}}\leq 0.43$.

\begin{table}[bt]
  \begin{center}\begin{tabular*}{\textwidth}{@{\extracolsep{\fill}}c|rrrr|rrrr|rrrr}
    \hline
    \hline
    Scenario\rule{0mm}{2.5em}
    & \multicolumn{4}{c|}{\shortstack{Hadronic\\topology}}
    & \multicolumn{4}{c|}{\shortstack{Semi-leptonic\\topology}}
    & \multicolumn{4}{c}{\shortstack{Combined\\topologies}} \\
    & obs. & $-1\sigma$ & exp. & $+1\sigma$
    & obs. & $-1\sigma$ & exp. & $+1\sigma$
    & obs. & $-1\sigma$ & exp. & $+1\sigma$ \\
    \hline
    $SVT$ &      1218 & 1268 & 1180 & 1097 &      1315 & 1406 & 1301 & 1203 &      1402 & 1468 & 1366 & 1264 \\
    $S$   &       577 &  604 &  556 &  520 &       647 &  647 &  603 &  555 &       685 &  693 &  641 &  593 \\
    $V$   &       953 & 1003 &  933 &  863 &       997 & 1069 &  997 &  921 &      1073 & 1141 & 1068 &  980 \\
    $T$   &      1069 & 1117 & 1045 &  969 &      1124 & 1232 & 1142 & 1052 &      1204 & 1300 & 1210 & 1114 \\
    $a$   &       436 &  462 &  430 &  400 &       472 &  513 &  475 &  436 &       499 &  535 &  499 &  459 \\
    $V-a$ &       961 & 1009 &  941 &  877 &      1018 & 1093 & 1018 &  938 &      1093 & 1163 & 1083 &  998 \\
    $V+a$ &       935 &  983 &  915 &  851 &       988 & 1078 & 1002 &  921 &      1058 & 1130 & 1058 &  974 \\
    \hline
    \hline
  \end{tabular*}\end{center}
  \caption{Observed and expected 95\% CL lower limits on $\Lambda$~(GeV)
  for the different scenarios, assuming $m_\Xt=175$~GeV$/c^2$. The $\pm 
  1\sigma$ values around the expected median limit are also shown.
  \label{tab:limits}}
\end{table}

The evaluation of the limits was performed taking into account
systematic uncertainties, which affect the background estimation and the
signal efficiency. The stability of the sequential analysis was studied by
changing the cut values in the most relevant variables by typically
10\%. The maximum relative change in the limit was about 2\%. Different
parameterisations inside PYTHIA were used to study the dependence of the
efficiency on the hadronisation and fragmentation of the jets associated
to heavy quarks. The Lund symmetric fragmentation, the Bowler space-time
picture of string evolution and the Peterson/SLAC function were
considered\footnote{See~\cite{pythia57a,pythia57b} for more details.}. The maximum
relative change in the limit was about 2\%. The effect of PDF binning
and smoothing procedures was studied and the maximum relative change in
the limit was about 3\%. A similar study was performed for the
discriminant variables and the maximum relative change in the limit was
about 6\%.


\section{Conclusions\label{sec:conclusions}}

Single top quark production via contact interactions was searched for
using data collected by the DELPHI detector at centre-of-mass energies
ranging from 189~GeV to 209~GeV, corresponding to an integrated
luminosity of 598.1~pb${}^{-1}$. The coupling scenarios listed in
Table~\ref{tab:scenarios} were considered and a dedicated analysis was
developed. No evidence for a signal was found. Limits at 95\% confidence
level on the new physics energy scale $\Lambda$ were obtained and the
observed values for different scenarios range from 499~GeV to 1402~GeV
(see Table~\ref{tab:limits}). The observed limit on the anomalous
coupling $\kappa_\XZ$, obtained from the conversion of scenario $a$
limit, is $\kappa_\XZ^{\mathrm{obs}}\leq 0.43$.

The L3 collaboration also searched for single $\Xt$ quark production via
contact interactions and the results~\cite{l3} are similar to those presented
here. The converted limit on the anomalous coupling $\kappa_\XZ$ agrees with
those presented by the four LEP collaborations~\cite{aleph00,aleph02,fcnc,l3,opal} in
the framework of Ref.~\cite{obraztsov}.


\subsection*{Acknowledgements}
\vskip 3 mm
We are greatly indebted to our technical 
collaborators, to the members of the CERN-SL Division for the excellent 
performance of the LEP collider, and to the funding agencies for their
support in building and operating the DELPHI detector.\\
We acknowledge in particular the support of \\
Austrian Federal Ministry of Education, Science and Culture,
GZ 616.364/2-III/2a/98, \\
FNRS--FWO, Flanders Institute to encourage scientific and technological 
research in the industry (IWT) and Belgian Federal Office for Scientific, 
Technical and Cultural affairs (OSTC), Belgium, \\
FINEP, CNPq, CAPES, FUJB and FAPERJ, Brazil, \\
Ministry of Education of the Czech Republic, project LC527, \\
Academy of Sciences of the Czech Republic, project AV0Z10100502, \\
Commission of the European Communities (DG XII), \\
Direction des Sciences de la Mati$\grave{\mbox{\rm e}}$re, CEA, France, \\
Bundesministerium f$\ddot{\mbox{\rm u}}$r Bildung, Wissenschaft, Forschung 
und Technologie, Germany,\\
General Secretariat for Research and Technology, Greece, \\
National Science Foundation (NWO) and Foundation for Research on Matter (FOM),
The Netherlands, \\
Norwegian Research Council,  \\
State Committee for Scientific Research, Poland, SPUB-M/CERN/PO3/DZ296/2000,
SPUB-M/CERN/PO3/DZ297/2000, 2P03B 104 19 and 2P03B 69 23(2002-2004),\\
FCT - Funda\c{c}\~ao para a Ci\^encia e Tecnologia, Portugal, \\
Vedecka grantova agentura MS SR, Slovakia, Nr. 95/5195/134, \\
Ministry of Science and Technology of the Republic of Slovenia, \\
CICYT, Spain, AEN99-0950 and AEN99-0761,  \\
The Swedish Research Council,      \\
The Science and Technology Facilities Council, UK, \\
Department of Energy, USA, DE-FG02-01ER41155, \\
EEC RTN contract HPRN-CT-00292-2002. \\




\end{document}